\documentclass[fleqn,usenatbib]{mnras}

\usepackage{newtxtext,newtxmath}

\usepackage[T1]{fontenc}

\DeclareRobustCommand{\VAN}[3]{#2}
\let\VANthebibliography\thebibliography
\def\thebibliography{\DeclareRobustCommand{\VAN}[3]{##3}\VANthebibliography}

\usepackage{graphicx}	
\usepackage{amsmath}	
\usepackage{cancel}
\usepackage[normalem]{ulem}

\title[The impact of disk outflows on SGRB structure]{The impact of disk outflows on the structure of short GRB jets at large scales}

\author[Urrutia, Janiuk \& Nouri]{
Gerardo Urrutia$^{1}$\thanks{E-mail: gurrutia@cft.edu.pl},
Agnieszka Janiuk$^{1}$, and
Fatemeh Hossein Nouri$^{1,2}$
\\
$^{1}$Center for Theoretical Physics, Polish Academy of Sciences, Al. Lotnikow 32/46, 02-668 Warsaw, Poland \\
$^{2}$ INFN, Sezione di Milano-Bicocca, Piazza della Scienza 3, I-20126 Milano, Italy
}

\date{Accepted XXX. Received YYY; in original form ZZZ}

\pubyear{2025}

\begin{document}
\label{firstpage}
\pagerange{\pageref{firstpage}--\pageref{lastpage}}
\maketitle


\begin{abstract}
Short Gamma-Ray Bursts (GRBs) are known to be associated with binary neutron star (NSNS) or black hole-neutron star (BHNS) mergers. The detection of gravitational wave and its associated electromagnetic counterparts GW/GRB 170817A has shown that interactions between relativistic jets and mildly relativistic ejecta influence observed radiation. Previous studies simulated a uniform jet propagating through a homologously expanding wind, however, jets and disk outflows are launched together during accretion, making the interaction more complex.  We investigate how the disk wind impacts jet propagation at distances $r\sim 10^8
\,-\,10^{11}~$cm. We are using two-dimensional special relativistic hydrodynamical simulations. As initial conditions, we remap the outflows from general relativistic magnetohydrodynamical simulations of BH accretion disks that represent post-merger NSNS or BHNS remnants. We account for wind stratification and r-process nucleosynthesis, which alter the pressure profile from that of an ideal gas in the initial conditions. We found that a) self-consistent wind pressure leads to significant changes in the jet collimation and cocoon expansion; b) the angular structure of thermal and kinetic energy components in the jets, cocoons, and winds differ with respect to simple homologous models; c) the temporal evolution of the structure reveals conversion of thermal to kinetic energy being different for each component in the system (jet, cocoon, and wind); d) dynamical ejecta alters the interaction between jets and disk winds. Our results show that the jet and cocoon structure is shaped by the accretion disk wind that alters the effect of dynamical ejecta and may have an impact on the observed afterglow emission.
\end{abstract}


\maketitle


\section{Introduction}

Short gamma-ray bursts (SGRBs) are intense flashes of gamma-rays lasting $t \lesssim 2$ s followed by a broad-band, long-lasting, afterglow radiation \citep{Frailetal1997,Metzger1997,Gehrels2005Natur,Bloom2006,NAKAR2007,Berger2014,kumar15}. 
These events presumably originate from binary neutron star, or black hole - neutron star mergers, as postulated by \cite {eichler89} and \cite {paczynski91}.
Other scenarios, including magnetars, or accretion induced neutron star collapse \citep{Thompson1995,Qin1998,Levan2006magnetar,Metzger2008,berger2011}, but they do not change drastically the main picture.
Once the central engine is created (a Kerr black hole surrounded by an accretion disk), the rotational energy of the black hole is transformed to the jet power, and its bulk kinetic energy, through the electromagnetic coupling \citep{bz77}. Alternatively, annihilation of neutrino-antineutrino pairs emitted from the accretion disk may be a source of power to the GRB jets \citep{Goodman1987,Ruffert1997,Popham1999,rosswogRamirezRuiz2002,Aloy2005,Liu2015,Zalamea2011}. The emission of SGRBs is further affected by the propagation of powerful (up to $E\sim 10^{52}$~erg) relativistic jets into a post neutron star merger environment.

From compact binary NSNS or BHNS mergers, a second electromagnetic counterpart, namely the kilonova, becomes observable. It is powered by nucleosynthesis occurring within the neutron-rich outflows after the merger \citep{Li-Paczynski1998,Kulkarni2005,Rosswog_2005,Metzger2010,Roberts_2011,Goriely_2011,Metzger_2012,radice2016,Rosswog2017,Radice_2018,Radice_2018c,radice2018b,Cowan2021,Perego2021book,Arcones2023}. 
The most prominent example is the detection of GW 170817 \citep{abbot2017NSmerger}, which not only provided detailed observations of electromagnetic counterparts, but also presented significant challenges for modeling. For instance, the unusual rise observed in the afterglow light curves of the \emph{off-axis} GRB, emphasized the necessity for meticulous treatment of this component in modelling \citep[e.g.,][]{lazzati18,MooleyNature2018}.

In addition to the relativistic jets, two broad sources of ejecta characterize the NSNS and BHNS mergers. The first is the dynamical ejecta generated from the NS tidal tail on a typically millisecond time scale. This ejecta is normally dominated by highly neutron-rich matter and is believed to be the origin of long-lasting near-infrared (red) kilonova~\citep{Metzger_2012,Hotokezaka-2013PhRvD,Grossman-2014,Tanaka-2014}. 
The viscous driven outflows originate in the remnant disk, operating on longer timescales ($\sim$ seconds), provide a second important source of ejecta. The disk wind ejecta contains less neutron-rich, lanthanide-free matter, causing the early optical-wavelength (blue) emission~\citep{Perego-2014,Kasen:2015,Tanaka:2017,Metzger-2017LRR,Fujibayashi-2018ApJ}.

The structure of the jet (energy and velocity distribution) plays an important role in understanding the atypical behaviour of the light curves \citep[e.g.,][]{granot2018,lazzati18,MooleyNature2018,Beniamini2020,Gill2019,Salafia2023structuredjets}. After its launching, the jet acquired its structure at early times $t \lesssim 0.1~$s due to the influence of the central engine properties such as the black hole spin, the magnetic field strength, and the strong disk winds evolution \citep[e.g.,][]{Aloy2005,Kathirgamaraju2019,James_2022,JaniukJames2022}. As long as the jet is evolving into a post-merger environment, the dynamics of the jet is influenced by ongoing interactions with this environment. In consequence, its structure can be modified far from the central engine \citep[e.g.,][]{lazzati18,Lazzati_2021_mergerEjecta,Hamidani2020,Hamidani&IOKAexpanding,Murguia_Berthier_2021}.

The interaction of the jet with the post-merger environment has been intensively studied with numerical simulations. This environment has been mainly described in two approaches. The first is to assume as a spherical wind expanding homologously and parametrized by its mass loss rate $\dot{M}_{\rm wind}$ and velocity \citep{Murguia2014,Nagakura_2014,Hamidani2020,Hamidani&IOKAexpanding,Urrutia2021ShortGRBS,Nativi2022,GottliebNakarExpandingMedia2021,Hamidani2023,Mpisketzis2024}. Alternatively, the wind can also be described by homologous toroidal winds, whose angular profiles of density $\rho_{\rm wind} (r,\theta)$ and velocity $v_{\rm wind} (r,\theta)$ were extracted from neutrino driven or strongly magnetized winds \citep{Aloy2005,Murguia-Berthier2017,Nativi2020,Murguia_Berthier_2021}. In addition to the wind characteristics, the dynamics of weakly magnetized jets has been explored \citep[][]{Nathanail2020,GottliebNakarExpandingMedia2021} and the impact of the magnetized environment \citep[e.g.,][]{GarciaGarcia2023}. In all these works, a wide range of the wind and jet properties were explored.

The second approach to study the jet-ejecta interaction, is to import the post-merger environment from simulations that evolve the binary NSNS coalescence until the creation of hypermassive neutron star (HMNS) \citep[e.g.,][]{Ciolfi2017,Ciolfi2019}. The results are remapped in a new numerical setup and the hydrodynamical evolution of the jet is analyzed \citep[e.g.,][]{Lazzati_2021_mergerEjecta,pavan2021}. The post-merger environment is not only described by density and velocity distributions. In addition, it presents turbulence and geometrical evolution of magnetic fields \citep{combi2023jets,Pavan2023,Pais_2023,Pais_2024}. However, in these works, the collapse of HMNS to a black hole was followed analytically, and the accretion disk formation was not considered. On the other hand, the evolution of accretion disk and jet from small to large scales has been solved, respectively, for NSNS and BHNS merger by \citet{gottlieb2022c,Gottlieb2023-BHNS}, showing that wind component impacts the large-scale evolution of the jet. These studies show that both dynamical ejecta and disk winds can significantly influence the short GRB jet's breakout and propagation.

The postmerger accretion disk's characteristics depend on the compact object merger parameters such as mass ratio, neutron star's equation of state and BH spin (in the case of BHNS)~\citep{Lovelace-2013,Foucart-2013,Kruger-2020}. Magnetic field and neutrino particles are the main rulers of the postmerger disk evolution. In addition to launching jets, magnetic field is responsible for heating the plasma~\citep{Sano-2004,Nouri-2018} and emergence of the magnetically-driven winds~\citep{Fernandez-2019,Janiuk2019,Fahlman2022}. They transport the angular momentum outwards, due to magnetorotational instabilty (MRI)~\citep[e.g.,][]{Balbus-2002}. The postmerger disks can be partially transparent to neutrinos i.e., the thermal energy can be lost by neutrino radiations \citep[e.g.,][]{Deaton2013,Nouri-2018}. Moreover, neutrinos provide mechanisms for neutrino viscosity~\citep{Guilet-2015,Guilet-2017} and launching thermal winds through neutrino absorption~\citep{Perego-2014,Martin-2015}. Therefore, both magnetic field and neutrinos have contributions in the thermal evolution and launching neutron-rich outflows for the r-process nucleosynthesis. It is important to note that the outflows properties (such as velocity, composition, temperature and opacity) are not only affected by the launching mechanism in the accretion disk, but also continuously change as a result of the heat released by the r-process nucleosynthesis~\citep{Lattimer1974,eichler89,Woosley-1992,Ruffert1997,Freiburghaus_1999,Rosswog_2005,Metzger2010,Roberts_2011,Goriely_2011,radice2016,Kasen2017,Radice_2018,Radice_2018c,Tanaka:2017}~\citep[reviews][]{Cowan2021,Perego2021book,Arcones2023}, hence in a realistic scenario, the ejecta's properties are more complex than a simple homologously expanding wind.

In this work, we study how the outflow from the post-merger disk impacts the propagation of the jet. In particular, we focus on two main progenitor types: the outflows originating from the central engines formed after the NSNS and BHNS mergers. The accretion disk wind profiles were obtained from numerical general relativistic (GR MHD) simulations in Kerr metric performed by \citet{Nouri_2023}, and then remapped onto a new setup to follow the large-scale jet evolution at late times.  In our analysis, we distinguish each component of the post-merger environment to emphasize the influence of the disk wind outflow on the jet properties. We also highlight the differences in comparison to the impact of a homogeneous post-merger environment.

This paper is structured as follows: In Section \ref{sec:methods} we describe our physical assumptions, the numerical setup, and how the disk wind is collected. In Section \ref{sec:results} we present our results, and Section \ref{sec:discussion} is dedicated to discussing the contribution of the disk wind regarding a homogeneous environment. Finally, in Section \ref{sec:conclusions} we summarize our results.


\section{Methods}\label{sec:methods}

%
%

\subsection{Numerical Setup}
%

We perform two-dimensional simulations in Special Relativistic Hydrodynamics (SRHD) to study the interaction of the post-merger disk wind and the jet at large scales and its impact on its final structure. We use the Adaptive Mesh Refinement (AMR) \emph{Mezcal} code \citep{decolle12}, which employs a second-order solver for both space and time to solve the SRHD equations. Primitive variables density $\rho$, velocities $v_j$, and pressure $p$ are determined through the evolution of conservative variables $D=\Gamma \rho $, $m_j=Dh\Gamma v_j$, and $\tau=Dh\Gamma c^2 - p -Dc^2$, by solving the system of equations
\begin{equation}
    \frac{\partial \mathbf{U(w)}}{\partial t} + \frac{\partial \mathbf{F^i(w)}}{\partial x^i} = 0.
    \label{eqn:hydro_eqs}
\end{equation}
Here, the conservative components are $\mathbf{U(w)} = \left(D, m_j, \tau\right)$, the flux is $\mathbf{F^i(w)}=\left(Dv^i,m_jv^i+p\delta_j^i,\tau v^i + pv^i \right)$, and $\Gamma$ represents the Lorentz factor. For the flux calculation, the Harten, Lax and van Leer Contact (HLLC) Riemman Solver \citep{MignoneBodo2005} is employed. 
Additional details about the \emph{Mezcal} code can be found in \cite{decolle12}.

We follow the jet and wind propagation during $t_I=4$~s of integration time. The initial conditions are imposed through the primitive variables $(\rho,v_j,p)$ at an inner boundary located at $r_{\rm inj} = 3\times 10^{8}~$cm. We use a two-dimensional computational box in a spherical coordinate system. It is extended radially from $r_{\rm inj}$ to $1.2\times 10^{11}~$cm, while the angular polar direction $\theta$ covers from $0$ to $\pi$.

For the grid configuration, we employ a number of cells $N_r=1\times 10^{4}$ along $r$ direction, and $N_\theta = 100$ in $\theta$ direction. We use $n_l=4$ levels of refinement which, under our configuration, being $r_{\rm max}=1.2\times 10^{10}~$cm, $r_{\rm inj}=3\times 10^{8}~$cm and $\theta_{\rm max}=\pi / 2$, the maximum resolution at the smallest cells is $\Delta r= (r_{\rm max}-r_{\rm inj})/(N_r\cdot 2^{n_l -1}) \approx 1.49 \times 10^{6}\,$cm and $r_{\rm inj} \Delta \theta = r_{\rm inj}\theta_{\rm max}/(N_\theta \cdot 2^{n_l -1}) \approx 1.17 \times 10^{6}\,$cm. 

We filled the computational box with a low-density medium $\rho_a=10^{-5}~$g cm$^{-3}$ to avoid numerical errors during the evolution of our short GRB models.

\subsection{Physical scenario and assumptions}

\begin{figure}
    \centering    
\includegraphics[scale=0.3]{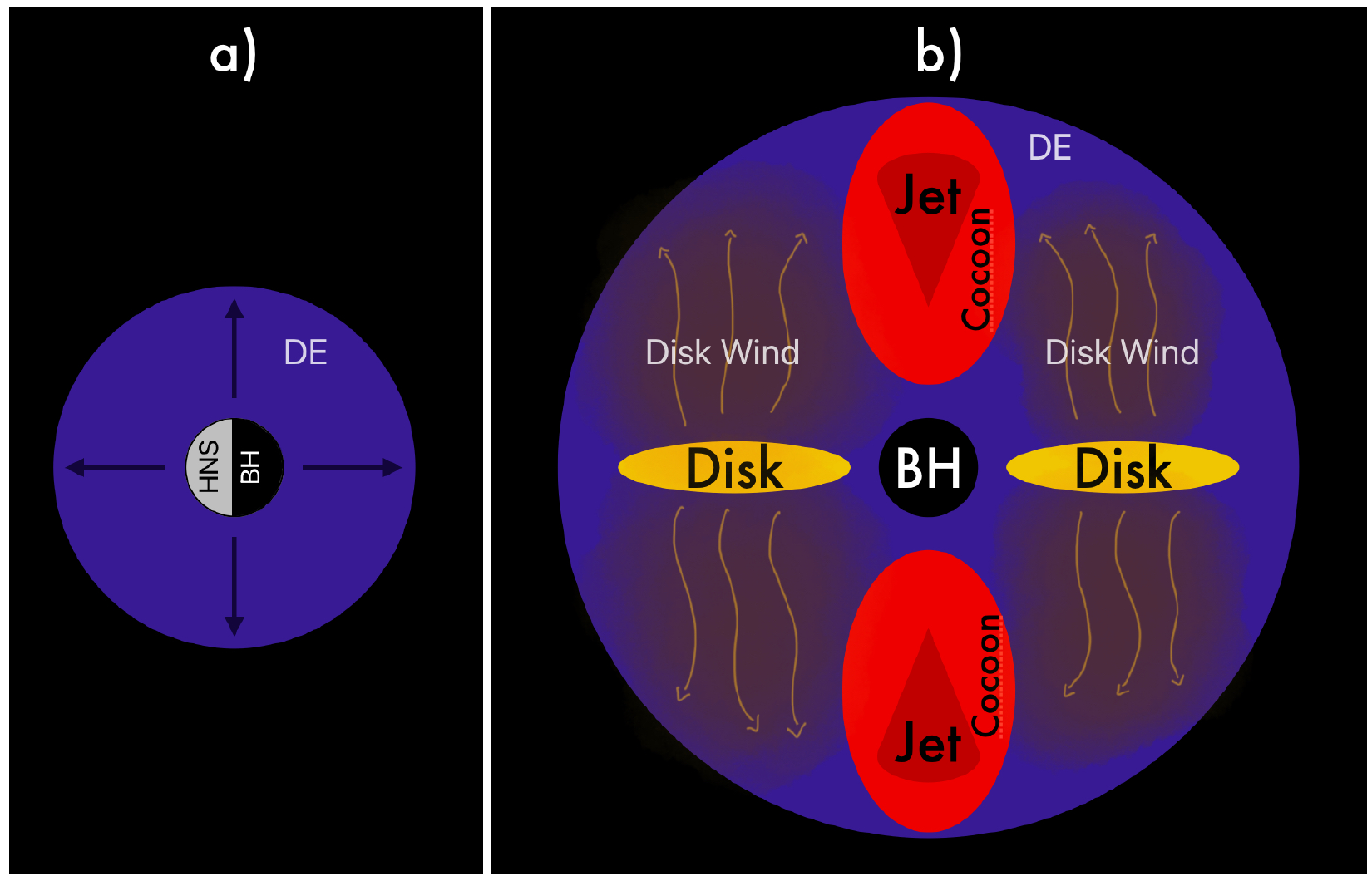}
    \caption{
    A non-scaled cartoon of a post-binary NSNS (or BHNS) merger. The post-merger ejecta is composed of the dynamical ejecta (DE) and the disk wind. In panel a), we assume that the DE is expelled uniformly until the collapse of the HMNS to a BH, assumed to occur during $t_{\rm collapse}= 0.1~$s. The DE suffers an initial expansion and fills the initial environment where the jet and disk wind will be propagating. For simplicity, we assume the same expansion time for ejections caused by the post-binary NSNS and BHNS merger. In panel b), we show the outflow components after $t>t_{\rm collapse}$ studied in this work. In this system, the DE, disk wind, and jet are interacting.}
       \label{fig:cartoon}
\end{figure}

To study jet and post-merger ejecta (EJ) interaction in a large-scale simulation, one needs to consider the combination of three main elements (Fig.~\ref{fig:cartoon}) assuming they are being generated during and after an NSNS or BHNS merger. These physical elements are:

1- Dynamical ejecta (DE): NSNS and BHNS mergers eject unbound matter through processes that depend on factors like total binary mass, mass ratio, and the equation of state. In NSNS mergers, ejecta masses typically range from $10^{-4}-10^{-2}M_{\odot}$ with velocities of $0.1-0.3$c~\citep{Hotokezaka-2013PhRvD}, while BHNS mergers can eject up to $0.1 M_{\odot}$ with similar speeds~\citep{Kyutoku-2015PhRvD}. In NSNS mergers, mass ejection occurs through two main processes: (1) ejecting shock-heated matter in a broad range of angular directions~\citep{Bauswein-2013ApJ,Hotokezaka-2013PhRvD}, and (2) ejecting spiral arms from tidal interactions during the merger, which expands outwards in the equatorial plane due to angular momentum transport~\citep{Bauswein-2013ApJ,Lehner-2016}. In cases of prompt collapse, mass ejection is suppressed due to immediate black hole formation. In BHNS mergers, tidal forces disrupt the neutron star, ejecting matter mainly in the equatorial plane, partially covering the azimuthal range, which may result in a stronger viewing angle dependence on kilonova emissions compared to NSNS mergers. 
In our model, this ejecta is artificially implemented considering a spherical geometry\footnote{3D simulations reveal that dynamical ejecta environments are asymmetric \citep[e.g.,][]{Rosswogetal2003,Rosswog_2005,radice2016,Radice_2018c,Ciolfi2019,Rosswog2023} and especially highly asymmetric in the BHNS post-merger case \citep[e.g.,][]{Kawaguchi2024}.}, which mimics the shock-driven ejecta during the merger.

2- Disk wind ejecta (WIND): In NSNS and some BHNS mergers, an accretion disk forms around the compact remnant object.  Disk masses are ranging between $0.01-0.3 M_{\odot}$, influenced by binary properties and the equation of state~\citep{Oechslin-Janka:2006}. Disk outflows launched over seconds or longer can exceed the mass of dynamical ejecta. Initially, in NSNS case, high accretion rates produce thermal neutrinos that drive mass loss~\citep{Popham1999}, but if a black hole forms promptly, this contribution is minimal~\citep{2013MNRAS.435..502F,Just:2015}. When a hypermassive neutron star remnant persists, strong neutrino-driven winds increase ejecta mass and alter its composition~\citep{Perego2014,Martin-2015,Richers:2015ApJ}. As the disk evolves, viscous processes in the form of MHD turbulence drive additional mass loss, leading to neutron-rich winds that contribute to r-process heavy element formation~\citep{Metzger2008,Metzger:2009MNRAS}. Late disk winds can be comparable to or larger than that in the dynamical ejecta with a more isotropic distribution~\citep{Wu:2016MNRAS,Fernandez-2019}. The magnetically-driven disk wind ejecta is implemented in our study from GRMHD simulations of magnetized accretion disks around spinning black holes.

3- Jets: Shortly after the merger, the remnant accretion disk around the 
black hole potentially powers relativistic jets associated with short-duration gamma-ray bursts. These jets can inject kinetic energy into the surrounding ejecta, forming a "cocoon" that accelerates its expansion. The cocoon may propagate through the formerly shocked medium (DE), possibly leading to a separate component of a 'macronova' \citep{NakarPiran_2017}. Moreover, the thermal energy from the jet can generate a separate electromagnetic transient known as cocoon emission. However, the luminosity of the cocoon depends on how efficiently the shocked jet material mixes with the surrounding baryon-rich ejecta that collimates the jet~\citep{Lazzati:2017MNRAS,Gottlieb:2018,Hamidani2023,Gutierrez2024}. 
In our simulations, uniform jets characterized by different parameters are initially injected into the environment consisting of both DE and WIND.

In our study, we employ a simplified description of DE. Our main interest is to study the properties of the jet interaction with the post-merger disk WIND. In the following sections, we describe in more detail the implementation of these elements. \\

As an additional test study, we perform an unrealistic scenario without DE to investigate only the propagation of jets and disk winds (see Table \ref{tab:list_models}), i.e., neglecting any environmental effects caused by DE. The jet and wind propagate in a constant background density medium $\rho_a$, which is significantly lower to prevent outflows from modifying their dynamics.

\subsubsection{The Dynamical Ejecta Implementation}

To consider the effect of DE, we launch a spherical ejection during $t\lesssim 0.1$~s, i.e., before the launching of the disk wind and jet. We consider two prescriptions of DE. For the first case, we assume the density is given by 
\begin{equation}
\rho_{\rm DE-1} = \frac{ \dot{M}}{4\pi r_{\rm inj}^2 v_{\rm ej} }\, ,
\label{eqn:dyn_floor}
\end{equation}
and a velocity of $v_{\rm ej}=0.1c$. This medium is extended until $r\approx 3\times 10^9$~cm. We assume that this DE has a mass loss rate of $\dot{M}=10^{-2}\,M_{\odot}$~s$^{-1}$ \citep[e.g.,][]{combi2023jets}. It is designated as ``DE-1'' in Table \ref{tab:list_models}.

A second case of DE, denoted as DE-2, is assumed as a spherical atmosphere \citep[e.g.,][]{Lazzati_2021_mergerEjecta} described by
\begin{equation}
\rho_{\rm DE-2} (r) = \rho_{\rm DE-1} e^{-r/r_{\rm inj}}, 
\label{eqn:sph_floor}
\end{equation}
where we fixed the same density as $\rho_{\rm DE-1}$ at the launching point. This is a static stratified environment, we assume that it was expanded until $r\approx 6.7\times 10^9$~cm before the jet launching. The objective of this scenario is to study the propagation of jets in a DE environment while mitigating the impact of disk wind effects. In Table \ref{tab:list_models}, this case is denoted as ``DE-2''.

\subsubsection{The Disk Wind Ejecta Implementation
}\label{sec:disk_implementation}

We map the disk wind outflows to provide an initial condition at the inner boundary of the jet injection, located at $r_{\rm inj}$. These outflows were extracted from the General Relativistic Magneto Hydrodynamic (GRMHD) simulation of 2D magnetized accretion disks performed by \citet{Nouri_2023}. Each of their simulation represented different possible scenarios of the post-merger disks, formed after compact objects collision. We selected here two cases, namely M2.65-0.1-a0.9 and M5.0-0.3-a0.9 from that study, which refers to NSNS and BHNS post-merger disks, respectively. The key parameters for defining these configurations used in \citet{Nouri_2023}, which include the black hole's spin, mass, and the accretion disk's mass at the initial time, can be found in Table \ref{tab:list_models}.

For our selected cases, the more massive BH, $M_{BH}=5 M_{\odot}$, and disk $M_{disk}=0.3 M_{\odot}$ in M5.0-0.3-a0.9 represent a possible configuration for a BH-disk remnants of a BHNS merger, while the less massive BH, $M_{BH}=2.65 M_{\odot}$ and disk $M_{disk}=0.1 M_{\odot}$ in M2.65-0.1-a0.9 represent a configuration for the remnants of an NSNS merger (see the discussion in~\cite{Kruger-2020,SiegelMetzger2018,Janiuk2019,fernandez2020} about possible disk and ejecta masses for different merger scenarios).
These systems were evolved for $t_{GR}=2\times 10^4$ in geometric units: $G=c=M=1$, which is equivalent to $\sim 0.35$s for NSNS and $\sim 0.4$s for BHNS in the physical units according to $t_g \equiv GM_{\rm BH}/c^3 $ unit conversion. These simulations were performed by the HARM\_COOL code \citep{2017ApJ...837...39J, Janiuk2019} which is the developed version of HARM originally presented by \cite{Gammie_2003}.

In HARM\_COOL version, the plasma is described by a realistic nuclear equation of state composed of free protons, neutrons, electron-positron pairs, and helium nuclei. The species are relativistic and have arbitrary degeneracy level. Their number densities are determined by the equilibrium condition assumed for weak nuclear reactions. The neutrino cooling effect was included in this study and computed for both the partially trapped neutrinos and for the optically thin regime. The details of the microphysics treatment for these simulations are given in ~\cite{Janiuk:2013}, ~\cite{2017ApJ...837...39J}, and \cite{Janiuk2019}.

\begin{figure}
    \centering
    \includegraphics[scale=0.4]{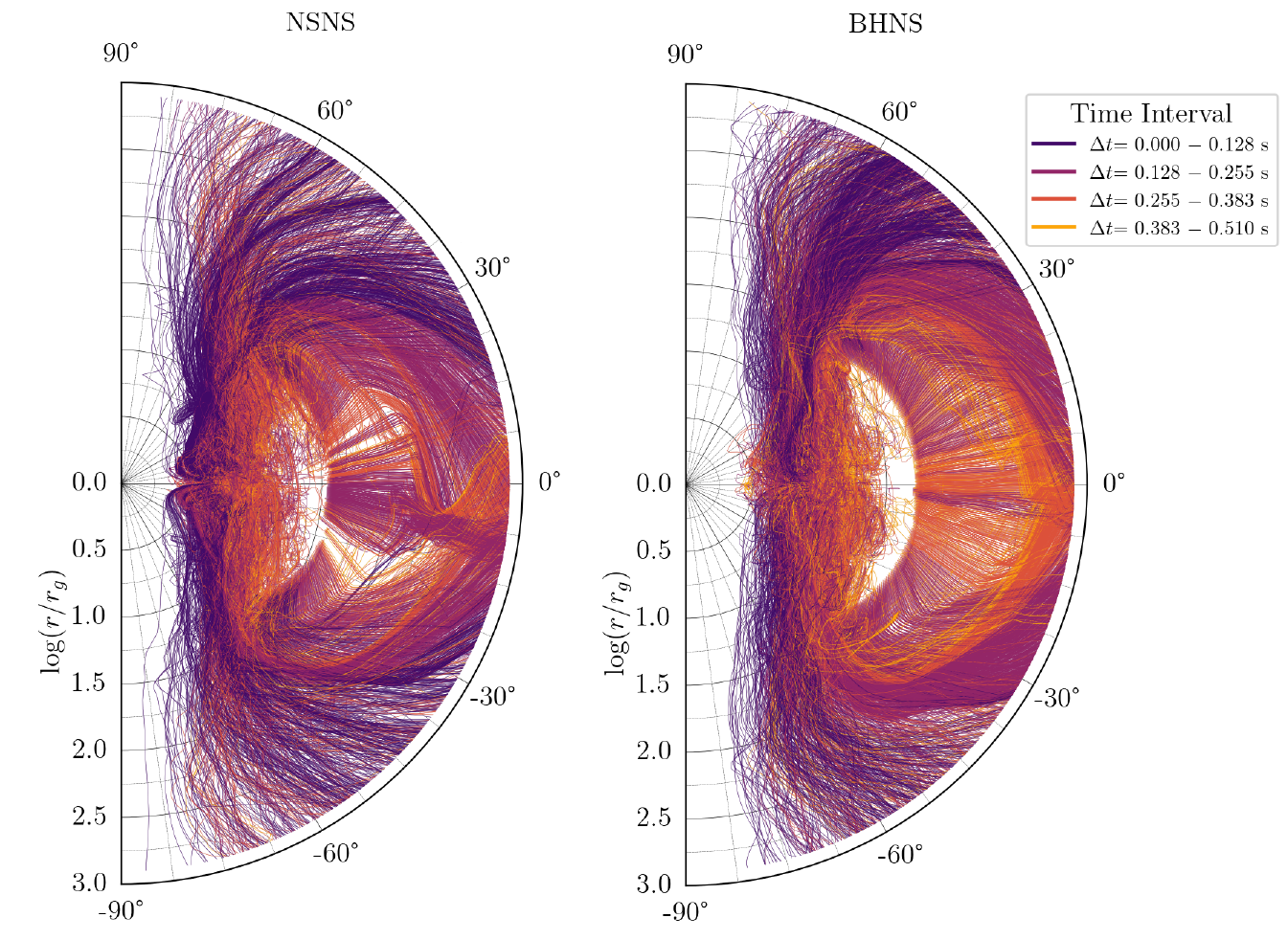}
    \caption{The trajectories of outflow tracers are projected onto a polar map with coordinates $(\log[r/r_g],\theta)$. Each color represents a specific time interval during which the outflow parcel spends reaching the external boundary of the GRMHD simulations, which corresponds to the internal boundary at $r_{\rm inj}$ in the SRHD simulations.}
    \label{fig:path-tracers}
\end{figure}

The evolution of the disk wind outflow is followed by a copious number of tracer particles. The tracer particle technique implemented in HARM\_COOL code has been described in detail in \cite{Janiuk2019}. In the initial state of the GRMHD simulation, the tracers are initialized in the grid, restricted to the densest parts of the flow (i.e., in the torus body). The code follows their trajectories in time, and saves those which leave the outer boundary. The two panels in Fig.~\ref{fig:path-tracers} refer to the disk wind outflows chosen from the work of \citet{Nouri_2023}.

We determine the composition of the ejecta by post-processing of the outflow tracers using the SkyNet nuclear reaction network \citep{Lippuner_2017}, which calculates the rate of change of abundance $\dot{Y}_i$ for each individual species $i$. The network employs the latest JINA REACLIB database. Upon the initial values of density and electron fraction on the tracers, the NSE solver determines the entropy per baryon. After that, the density temporal profile is read on tracers, while the tracer temperature evolves consistently to the expansion, accounting for nuclear heating. Hence, the temperature evolved in the network is influenced by the r-process nucleosynthesis, as well as is the nuclear pressure. We notice that this approach is still simplified with respect to the simulation where the hydrodynamics would be coupled to the nuclear network at each time-step. The latter can produce some differences in the abundances of specific isotopes, formed at late times \citep[e.g.,][]{Magistrelli2024}. This is however not relevant to our present study.

Notice that the previous work of \citet{Nouri_2023} presented the evolution of nuclear network for long times; in the present study, we limited the integration to the time which takes the tracers to reach the inner jet boundary at $r_{\rm inj}$ (individually, each of them touches the inner boundary at different moment, until
$t\lesssim 0.4~$s). At this particular point, we extract the gas pressure from each tracer processed by SkyNet. We employed an inversion of the Helmholtz equation using the code presented by \cite{TimmesArnett1999ApJS}.

\begin{figure}
    \centering
    \includegraphics[scale=0.42]{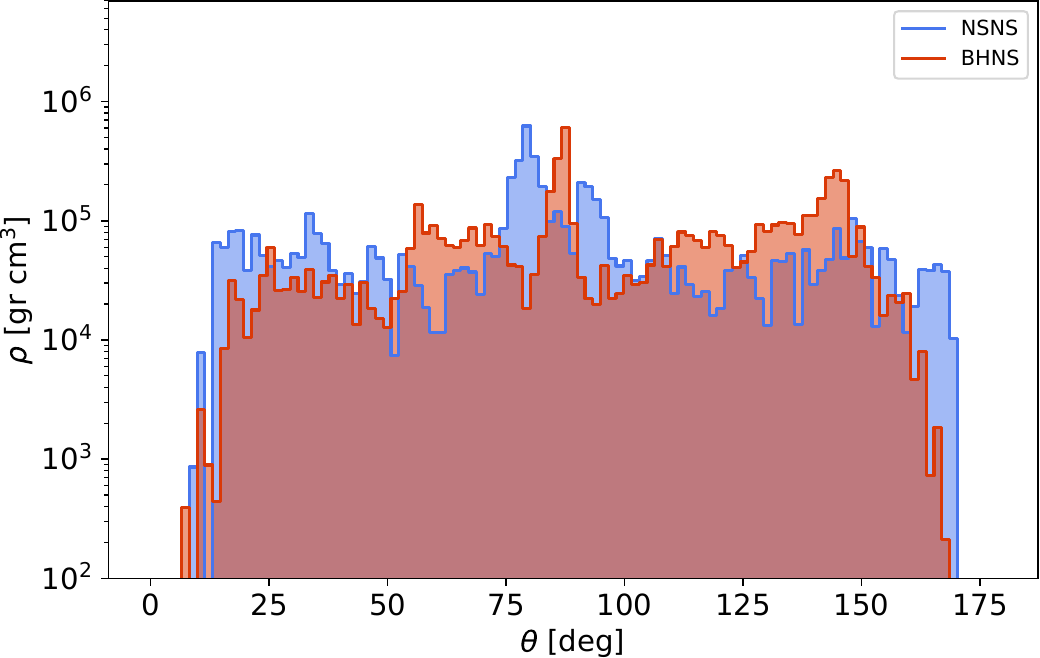}
    \includegraphics[scale=0.42]{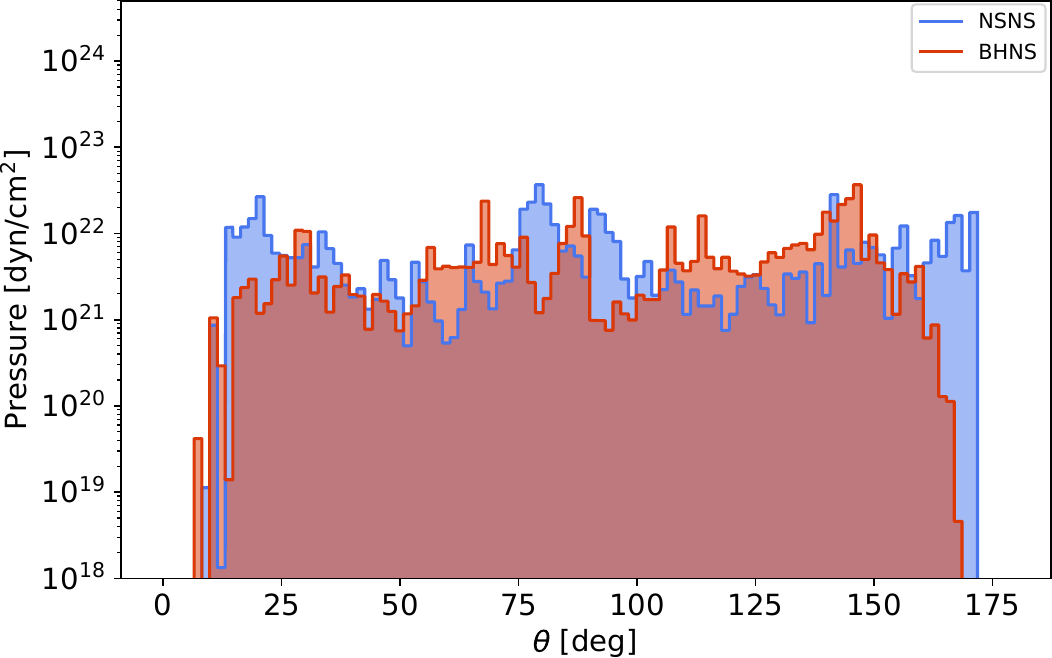}
    \includegraphics[scale=0.42]{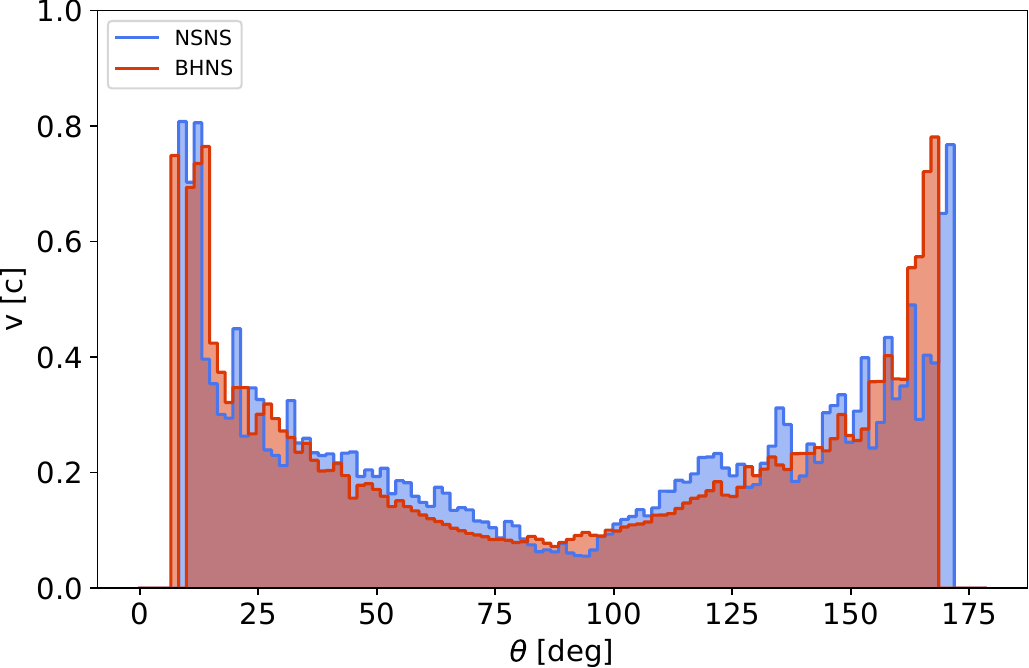}
    \includegraphics[scale=0.42]{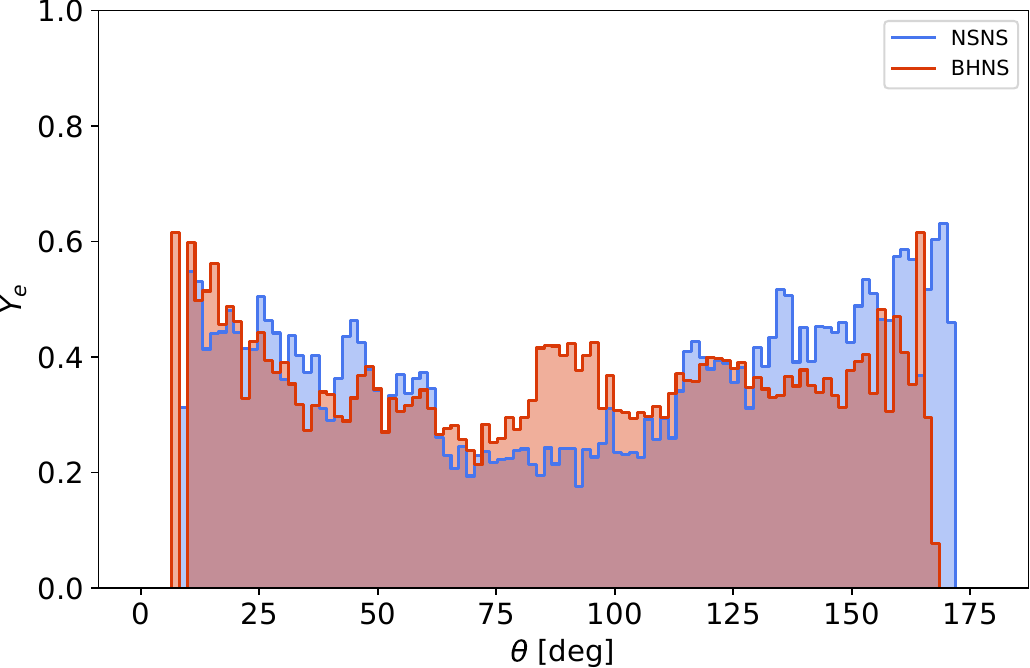}
    \caption{Snapshot at $t=0.3~$s of the outflow distribution. This data was obtained by collecting and averaging the quantities stored in tracer particles that reached the injection radius $r_{\rm inj}$. The values for pressure and electron fraction $Y_e$ were determined through the r-process.}
    \label{fig:histo_f}
\end{figure}

Snapshots of the primitive quantities $(\rho_w,v_w,p_w)$ distribution of the wind are shown in Fig.~\ref{fig:histo_f} as examples. Generally, there are common features shared by all the cases, for instance, the outflows are more massive on the equator, while the faster and less neutron-rich outflows are distributed on the poles.

\begin{figure}
    \centering
    \includegraphics[width=0.95\linewidth]{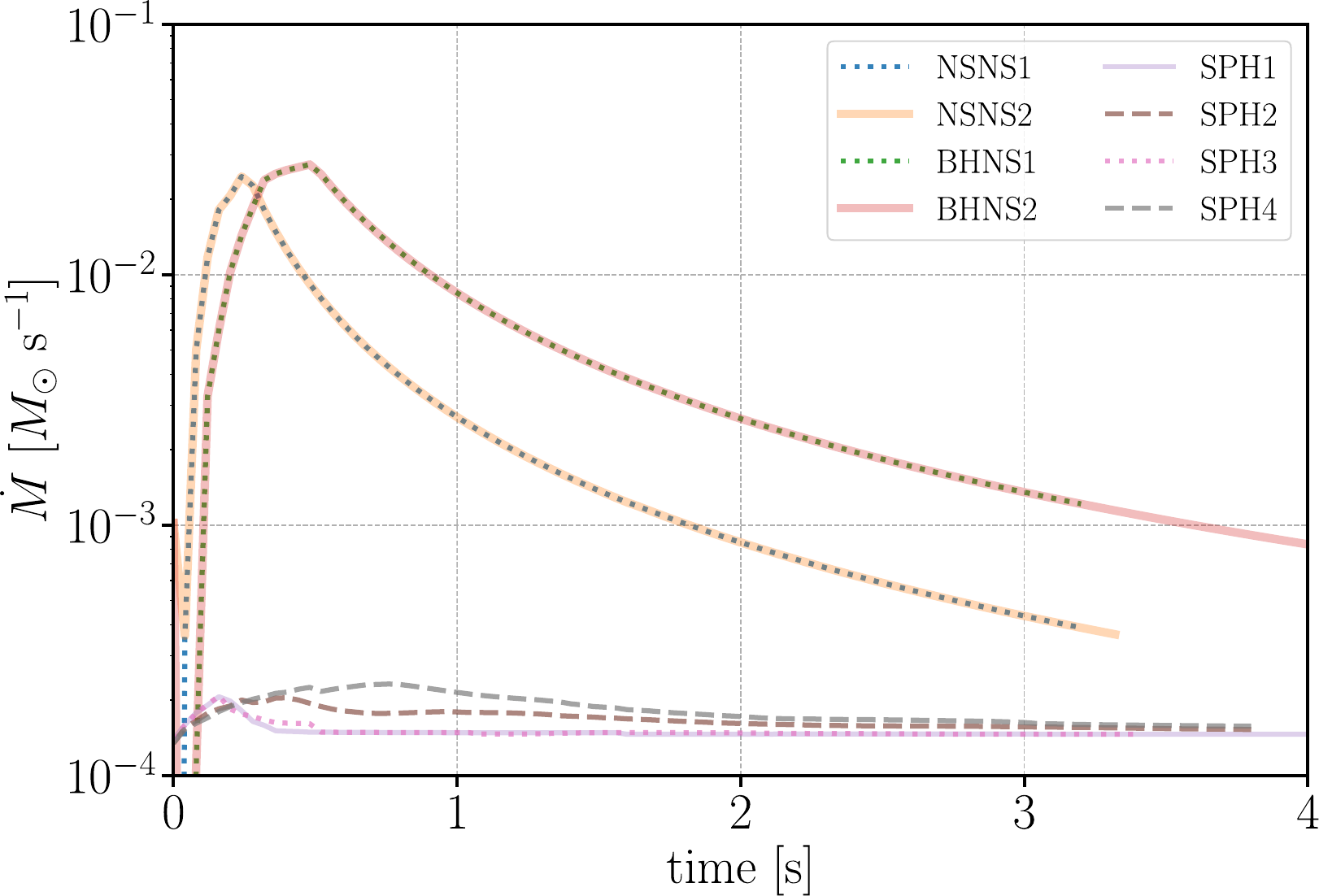}
    \caption{The total mass injection rate for different cases. The mass of the disk wind ejecta dominates over the dynamical ejecta in our initial conditions, as illustrated for (NSNS1,NSNS2) and (BHNS1,BHNS2) cases.}
    \label{fig:injection_mass}
\end{figure}

To assume a realistic physical timescale in our study, the integration time $t_{I}$ of large-scale SRHD simulations is greater than the integration time $t_{GR}$ of GRMHD simulations, where the disk wind ejecta has been originally extracted. After $t_{GR}$, we continue to inject the wind assuming the rate of $\dot{M}=\dot{M}_{\rm wind} (t/t_{GR})^{-5/3}$ for later times\footnote{Long-term simulations of accretion disks formed after binary mergers, for example, \citet{Fernandez-2019}, show that after 1 second the mass loss rate for outflow material is $\dot{M}_{\rm wind}\propto t^{-2.3}$, however, this slope could change due to the initial condition of the simulation.} \citep{Michel1988,Rees1988,Chevalier1989,LeeRamirezRuiz2007,Murguia2014,Murguia2017}. See Table \ref{tab:list_models} for the values of $\dot{M}_{\rm wind}$ corresponding to each case and Fig.~\ref{fig:injection_mass} for the evolution of the injection rate.

\subsubsection{The jet implementation}

Hereafter, all quantities referring to the jet will be denoted by the subscript $j$. The jet is characterized by its initial parameters: its opening angle $\theta_j$, which defines the angular region where the maximum energy is initially enclosed (or the jet core); its luminosity $L_j$, which accounts for magnetic, thermal, and kinetic energy contributions; the injection time $t_j$ which represents the mean lifetime, or the duration over which the energy is extracted from the central engine; the initial Lorentz factor $\Gamma_j$, which defines that the jest is relativistic, and the maximum Lorentz factor of the jet $\Gamma\infty$ achieved at large distances.


The jet parameters in our simulation were chosen to be consistent with the central engine in the GRMHD simulations. Following \citet{Nouri_2023}, we specify the jet opening angle as $\theta_j=15^\circ$, consistent with the mass distribution of the wind (Fig.~\ref{fig:histo_f}), specifically excluding the regions near the poles where wind is absent. The maximum Lorentz factor is assumed to be a fixed parameter $\Gamma_\infty=100$. 

Since we are following the central engine activity from two post-merger disk configurations, NSNS and BHNS, defined by their initial masses $m_d$ (see Section \ref{sec:disk_implementation}), the injection time is constrained by the average lifetime of the central engine \citep[e.g.,][]{1998ApJ...494L..53K,2002ApJ...577..893L,2008ApJ...675..519J} and approximated by
\begin{equation}
t_j = m_d/\langle \dot{M} \rangle\;.
\end{equation}
We found $t_j \sim 1.57$~s for NSNS merger scenario and $t_j \sim 1.07$~s for BHNS, respectively. The Lorentz factor $\Gamma_j$ was extracted directly from the final snapshot of the GRMHD simulation for both scenarios. The luminosity of the jet, $L_j$, was obtained by integrating the total energy density enclosed in the jet region $\theta \leq \theta_j$, as
\begin{equation}
L_j = 2\pi \int_0^{\theta_j} e_{\rm tot} v_r , r_{\rm inj}^2 , d\theta \, ,
\label{eqn:jet_lum}
\end{equation}
where,
\begin{equation}
e_{\rm tot} = \Gamma^2 h \rho c^2 - p - \Gamma\rho c^2\, ,
\label{eqn:total_en}
\end{equation}
with the specific enthalpy given by $h = 1 + \frac{\gamma}{\gamma-1} \frac{p}{\rho c^2}$, and $\gamma = 4/3$ being the adiabatic index of the gas. We are excluding the magnetic component in the total energy density since our SRHD simulations are hydrodynamical (the contribution of the magnetic field is described for example in \cite{GottliebNakarExpandingMedia2021,James_2022,JaniukJames2022,Pavan2023}). The list of the initial parameters of the jet is given in Table \ref{tab:list_models}.\\

The jet implementation in the \emph{Mezcal} code is through the primitive quantities $(\rho_j, v_j, p_j)$. The relation of these quantities with the initial parameters is given as follows. The density is related to the luminosity by
\begin{equation}
L_j = \Gamma_\infty \dot{M}_j c^2\, ,
\end{equation}
where $\dot{M}_j = \Gamma_j \rho_j \Delta V / \Delta t$ is the mass loss rate of the jet injected per unit volume and time \citep[e.g.,][]{urrutia22_3D}. By rearranging terms, the density of the jet is expressed as
\begin{equation}
\rho_j = \frac{L_j}{\Gamma_\infty \Gamma_j v_j c^2 \Delta S}\, ,
\label{eqn:den_high_press}
\end{equation}
being $\Delta S = 4 \pi \left( 1 - \cos \theta_j \right) r_{\rm inj}^2$. The maximum and initial Lorentz factor are related throughthe  jet's enthalpy $h_j$ by 
\begin{equation}
\Gamma_\infty = h_j \Gamma_j \, ,
\label{eqn:lor_fac_inf}
\end{equation}
where the pressure of the jet can be obtained by rearranging terms as
\begin{equation}
p_j = \frac{\rho_j c^2}{4} \left( \frac{\Gamma_\infty}{\Gamma_j} -1 \right)\, .
\label{eqn:high_jet_pressure}
\end{equation}
In summary, the initial conditions of the jet and wind are injected into an inner boundary located at $r_{\rm inj}$ and along $\theta \in [0, \pi]$. The jet region is filled as $(\rho, v, p) = (\rho_j, v_j, p_j)$ for $\theta \leq \theta_j$ and $\theta \geq \pi - \theta_j$ during $t\leq t_j$, while outside of the jet region the wind values are injected as $(\rho, v, p) = (\rho_w, v_w, p_w)$. 

\begin{table*}
\centering
\begin{tabular}{ |c||c|c|c|c|c|c|c|c|c||c|c|c|} 
 \hline
 Model & $L_j\times 10^{50}$ & $\theta_j$ & $\Gamma_j$ & $t_j$ & $M_{\rm BH}$ & $a$ & $M_{{\rm disk},0}$ & $M_{\rm wind}$ & $\dot{M}_{\rm wind}$ & EJ & $M_{\rm ini}^{\rm DE}$ & $r_{\rm ini}^{\rm DE}\times 10^9$\\ 
   & [erg~s$^{-1}$] & [deg] & & [s] & [$M_\odot$]  & & [$M_\odot$] & [$M_\odot$] & [$M_\odot$~s$^{-1}$] & & [$M_\odot$] & [cm]  \\ 
 \hline \hline
 NSNS1 & $1.4$ & $15^\circ$ & 7.4 & 1.57 & 2.65 & 0.9 & 0.1 & $7.7\times 10^{-4}$ & $3.3\times 10^{-2}$ &  WIND & -- &-- \\ 
 NSNS2 & $1.4$ & $15^\circ$ & 7.4 & 1.57 & 2.65 & 0.9 & 0.1 & $7.7\times 10^{-4}$ & $3.3\times 10^{-2}$ & DE-1 + WIND & $3.3\times 10^{-2}$ & $3.0$\\  
 SPH1 & $1.4$ & $15^\circ$ & 7.4 & 1.57 & -- & -- & -- & No disk wind & -- & DE-2  & $6.1\times 10^{-4}$ & $5.4$\\
 SPH2 & $1.4$ & $8^\circ$ & 7.4 & 1.57 & -- & -- & -- & No disk wind & -- & DE-2 & $6.1\times 10^{-4}$ &$5.4$ \\ 
 \hline \hline
 BHNS1 & $2.13$ & $15^\circ$ & 12 & 1.07 & 5 & 0.9 & 0.3 & $6.6\times 10^{-3}$ & $1.5\times 10^{-1}$ & WIND & --\\ 
 BHNS2 & $2.13$ & $15^\circ$ & 12 & 1.07 & 5 & 0.9 & 0.3 & $6.6\times 10^{-3}$ & $1.5\times 10^{-1}$ & DE-1 + WIND & $3.3\times 10^{-2}$&$3.0$\\  
 SPH3 & $2.13$ & $15^\circ$ & 12 & 1.07 & -- & -- & -- & No disk wind & -- & DE-2 & $6.1\times 10^{-4}$&$5.4$\\
 SPH4 & $2.13$ & $4.7^\circ$ & 12 & 1.07 & -- & -- & -- & No disk wind & -- & DE-2 & $6.1\times 10^{-4}$&$5.4$\\ 
 \hline
\end{tabular}
    \caption{The list of models performed in our study, detailing their main parameters and scenarios. The left side contains the jet parameters $(L_j,\theta_j,\Gamma_j,t_j)$ extracted from the GRMHD simulation data, whereas the central
    panel presents the configuration of the central engine $(M_{\rm BH},a,M_{{\rm disk},0})$ utilized in those simulations. The values of outflow mass were estimated by post-processing tracer outflows from GRMHD simulation are denoted by $(M_{\rm wind},\dot{M}_{\rm wind})$. The column denoted as ``EJ'' shows which components of the post-merger ejecta are initialized in our simulations. The label WIND refers to the disk wind, indicating whether it is initialized or not. The label DE-1 refers to the dynamical ejecta given by eq. \eqref{eqn:dyn_floor}, while DE-2 is given by eq. \eqref{eqn:sph_floor}. The initial parameters of the dynamical ejecta are the initial mass $M_{\rm ini}^{\rm DE}$ and the initial radius $r_{\rm ini}^{\rm DE}$. This table is divided into two sections: the upper section defines the post-binary NSNS merger scenario, and the lower section corresponds to the post-binary BHNS merger scenario. }

    \label{tab:list_models}
 \end{table*}

\subsection{The energy extraction for post-processing analysis}

We define the total energy contribution as the sum of the energy enclosed in all cells \citep[e.g.,][]{Urrutia2021ShortGRBS,urrutia22_3D,Nativi2022} 
\begin{equation}
    E_{\rm} = \int e_{\rm tot}\, dV\, .
    \label{eqn:tot_en}
\end{equation}
The total energy density $e_{\rm tot}$ given by eq. \eqref{eqn:total_en}, can be expressed by its components $e_{\rm tot}=e_{\rm kin}+e_{\rm th}$ \citep[e.g.,][]{Suzuki2022,urrutia22_3D}, respectively, the kinetic energy density 
\begin{equation}
    e_{\rm kin} =  \Gamma\left(\Gamma-1 \right)\rho c^2\, ,
    \label{eqn:kin_en}
\end{equation}
and, the thermal energy density
\begin{equation}
    e_{\rm th} =  p \left(4\Gamma^2 -1 \right)\, .
    \label{eqn:th_en}
\end{equation}
The total contribution of independent components can be obtained by the integration of $e_{\rm kin}$ or $e_{\rm th}$, instead of $e_{\rm tot}$ in eq. \eqref{eqn:tot_en}. 

In addition, we divide the contributions of each component of the relativistic and sub-relativistic material, by a selection of how fast it moves. The jet is defined as the faster material, 
\begin{equation}
    E_{\rm jet} \equiv E(\Gamma \ge \Gamma_j), 
    \label{eqn:tot_en_jet}
\end{equation}
and the cocoon as an intermediate values
\begin{equation}
        E_{\rm cocoon} \equiv E (2 \leq \Gamma  < \Gamma_j).
        \label{eqn:tot_en_cocoon}
\end{equation}
The contribution of the wind requires and spatial treatment since the velocities are sub-relativistic. Following the definition of total energy in eq. \eqref{eqn:total_en}, the Lorentz factor in the kinetic contribution eq. \eqref{eqn:kin_en}, i.e., has influence $f(\Gamma)\equiv \Gamma(\Gamma-1)$. For a smoothed transition close to $\Gamma \sim 1$, we fix a minimum value $\Gamma_{\rm min}=1.1$ that corresponds to a minimum wind velocity of $\beta \sim 0.1c$ and we took the Taylor expansion of $f(\Gamma)$ close to $\Gamma_{\rm min}$. The kinetic contribution of the wind is expressed as
\begin{equation}
    e_{\rm w,kin} =   \frac{1}{2} D \, \beta^2 c^2\, ,
\end{equation}
where the rest mass in this limit is $D\equiv f(\Gamma)\,\rho$. The thermal energy remains unaffected close to $\Gamma_{\rm min}$ because the values $e_{\rm th} \sim 3p$. The energy of the wind is selected through a threshold of velocities
\begin{equation}
    E_{\rm wind} \equiv E (0.1 \leq \beta < 0.4)\, 
    \label{eqn:tot_en_wind}
\end{equation}
which represents the contribution of the slower material in the system. A similar selection with 4-velocity is made by \cite{gottlieb2017,Suzuki2022} and trough the asymptotic Lorentz factor by \cite{Lazzati_2021_mergerEjecta,HamidaniIoka2023b,GarciaGarcia2023}.

\section{Results}\label{sec:results}

In this section, we present the results of our simulations. First, we describe the interaction between the jet, the disk wind, and the DE. Second, we discuss two additional models that show the jet interacting without the disk's wind contribution. Finally, we present the evolution of the energy components extracted from the jet, cocoon, and wind.

\subsection{Jet and wind interaction}

\begin{figure*}
    \centering
    \includegraphics[scale=1.15]{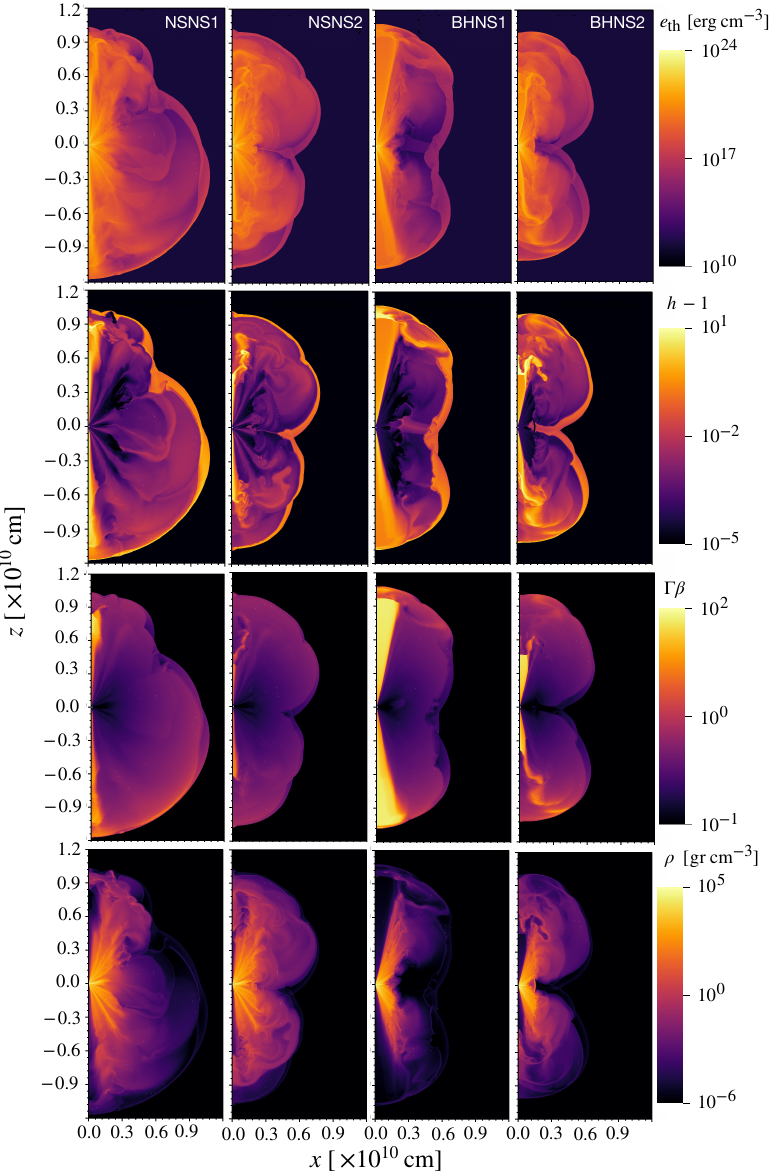}
    \caption{Color maps at $t=0.4~$s. Columns depict the models NSNS1, NSNS2, BHNS1, and BHNS2 (as detailed in Table \ref{tab:list_models}). Rows showcase density maps of thermal energy $e_{\rm th}$, enthalpy $h-1$, four-velocity $\Gamma\beta$, and density $\rho$, respectively from top to bottom. }
    \label{fig:color_maps}
\end{figure*}

In Fig.~\ref{fig:color_maps}, the columns display the models summarized in Table~\ref{tab:list_models}. They are categorized into two groups based on the progenitor, from either binary NSNS or BHNS mergers. The numerical labels, '$1$' or '$2$', indicate whether the jet and disk wind are interacting with an initial DE. Each row corresponds to a physical quantity: thermal energy $e_{\rm th}$, enthalpy $h-1=4p/(\rho c^2)$, the four-velocity $u=\Gamma\beta$, and density $\rho$, respectively displayed at $t\sim 0.4~$s.

During the early phase of propagation (e.g., $t\lesssim 0.4~$s), the presence of DE plays an important role in mediating the interaction between the jet and disk wind. This interaction can be observed in models NSNS2 and BHNS2 in which the jet and wind are immersed in dynamical ejecta. The shock front of the jet and wind modifies the propagation velocity due to mass accumulation. The pressure balance between the jet, cocoon and wind is altered due to the interaction with a dense, stratified environment, then the jet is collimated. In contrast, in the models NSNS1 and BHNS1 where the jets propagate only in a low background density environment, the distribution of the material is dominated by the initial velocity distribution and it is not modified by a dense environment. In consequence, the initial asymmetric distribution at the south pole of NSNS1 is preserved.

The thermal energy density map (1st row) provides a qualitative visualization of the interaction between the jet and the disk wind. In the NSNS1 and BHNS1 models, the thermal energy is significantly concentrated near the poles, with lower concentrations at the equator. The distribution of thermal energy becomes more uniform in the NSNS2 and BHNS2 models.

The enthalpy map (2nd row) highlights regions where thermal energy dominates over rest-mass energy. These heated regions are primarily concentrated in the zones of interaction, including the jet-cocoon balance, the cocoon-wind transition, and the shock at the boundary with the external low-density gas. In particular, this external shock creates a shell of swept-up material, which is more pronounced for the NSNS1 and BHNS1 models.

The velocity color map (3rd row) provides a quantitative structure of the jet ($u\sim 0.9\Gamma_j$), cocoon ($0.9\Gamma_j \gtrsim u \gtrsim 2$), and disk wind ($u \lesssim 2$). Darker regions correspond to material with $\beta \leq 0.1$. In both NSNS1 and NSNS2 models, the disk wind significantly impacts the jet, leading to its collimation. The south pole of NSNS1 exhibits a whiplash effect, resulting in the formation of a faster shell. The BHNS models illustrate the propagation of jets injected with high Lorentz factors, as detailed in Table \ref{tab:list_models}. The jet channel continues to expand in BHNS1, while it remains collimated in BHNS2. Both models produce fast external shells. The disk wind velocity in all models remains sub-relativistic due to the denser outflows.

The density maps (4th row) illustrate the asymmetrical distribution of ejected material. Notably, denser material accumulates toward the inner regions. This behaviour can be attributed to heavy material tending to maintain lower velocities. A substantial fluid, characterized by high-density concentrations, exhibits less diffusive propagation. For instance, the fluid in the BHNS scenario, imposed as a massive ejection (see Table \ref{tab:list_models}) is more diffusive than in the NSNS scenario, which shows a higher degree of mixing.

\subsection{Jet propagation in spherical wind}

\begin{figure}
    \centering
    \includegraphics[scale=0.4]{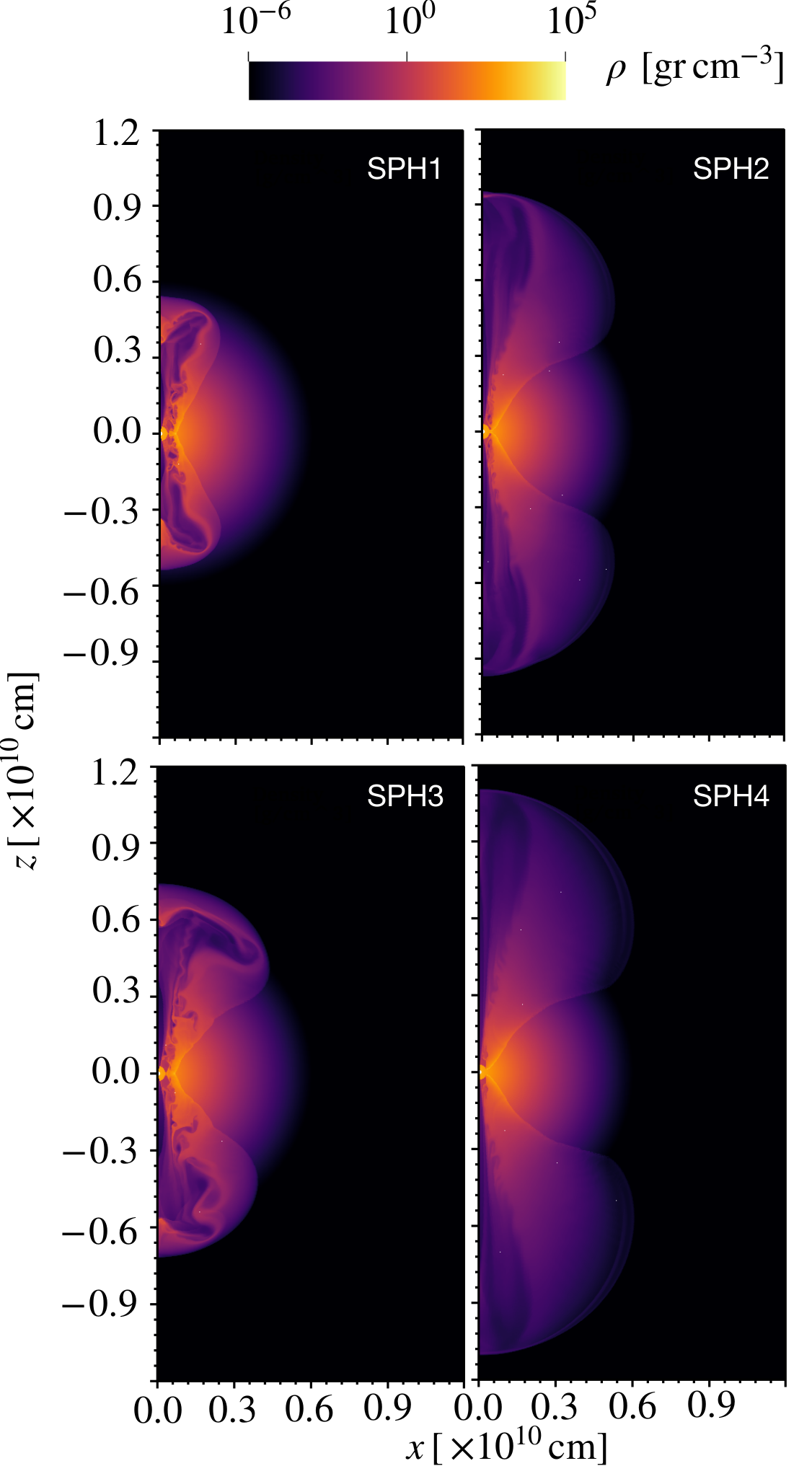}
    \caption{Density maps at $t=0.4~$s of jets propagating into a spherical atmosphere without wind contribution. In the upper panel, the models SPH1 and SPH2 belong to the group post binary NSNS merger, while in the lower panel, SPH3 and SPH4 belong to the group post binary BHNS merger as we shown in Table \ref{tab:list_models}.}
    \label{fig:sphmodels}
\end{figure}

For a comparative analysis, we performed four additional jet models propagating within a simplified spherical atmosphere, which is defined by eq.~\eqref{eqn:sph_floor} and labelled as DE-2, excluding any interactions with the disk wind. We studied two types of jets, collimated \citep[$\theta_j < 1/\Gamma_j$, e.g.,][]{Bromberg2011} and non-collimated ($\theta_j = 15^\circ$). The jet models SPH1 and SPH2 correspond to the NSNS models, while SPH3 and SPH4 correspond to the BHNS models (see Table \ref{tab:list_models}). The maps presented in  Fig.~\ref{fig:sphmodels} show a slow propagation for non-collimated jets, while the initial collimated cases present the formation of re-collimation shocks, and the cocoon expands according to atmospheric density declines for long distances. Jet models NSNS and BHNS (Fig.~\ref{fig:color_maps}) were implemented as non-collimated, and it is observed that the disk wind collimates the jet. 

\subsection{Chemical composition of the wind}

Studying the chemical composition of the disk wind and its interaction with the jet and DE is crucial for predicting kilonova light curves \citep[e.g.,][]{NakarPiran_2017,gottlieb2017,Hamidani2023,Hamidani_2024lateactivity,Hamidani_2024GRB211211}. In a merger scenario, the less neutron-rich and more transparent disk wind is surrounded by the more opaque and neutron-rich DE produced during the merger~\citep{Metzger:2010,Grossman-2014,Kawaguchi:2016}. Based on the study by~\cite{Kasen:2015} using a radiative transfer code, the presence of even a small amount of overlying, neutron-rich DE ($10^{-4}M_{\odot}$) acts as a `lanthanide-curtain' blocking the optical wind emission from certain viewing angles. The \emph{Mezcal} code \citep{decolle12}, employed in the present study, is not formulated for radiative transfer and composition evolution. Therefore, we limit our discussion to the qualitative geometrical evolution of the wind, which contains matter with varying electron fractions, while the composition of DE has been neglected.

\begin{figure*}
    \centering
    \includegraphics[scale=0.95]{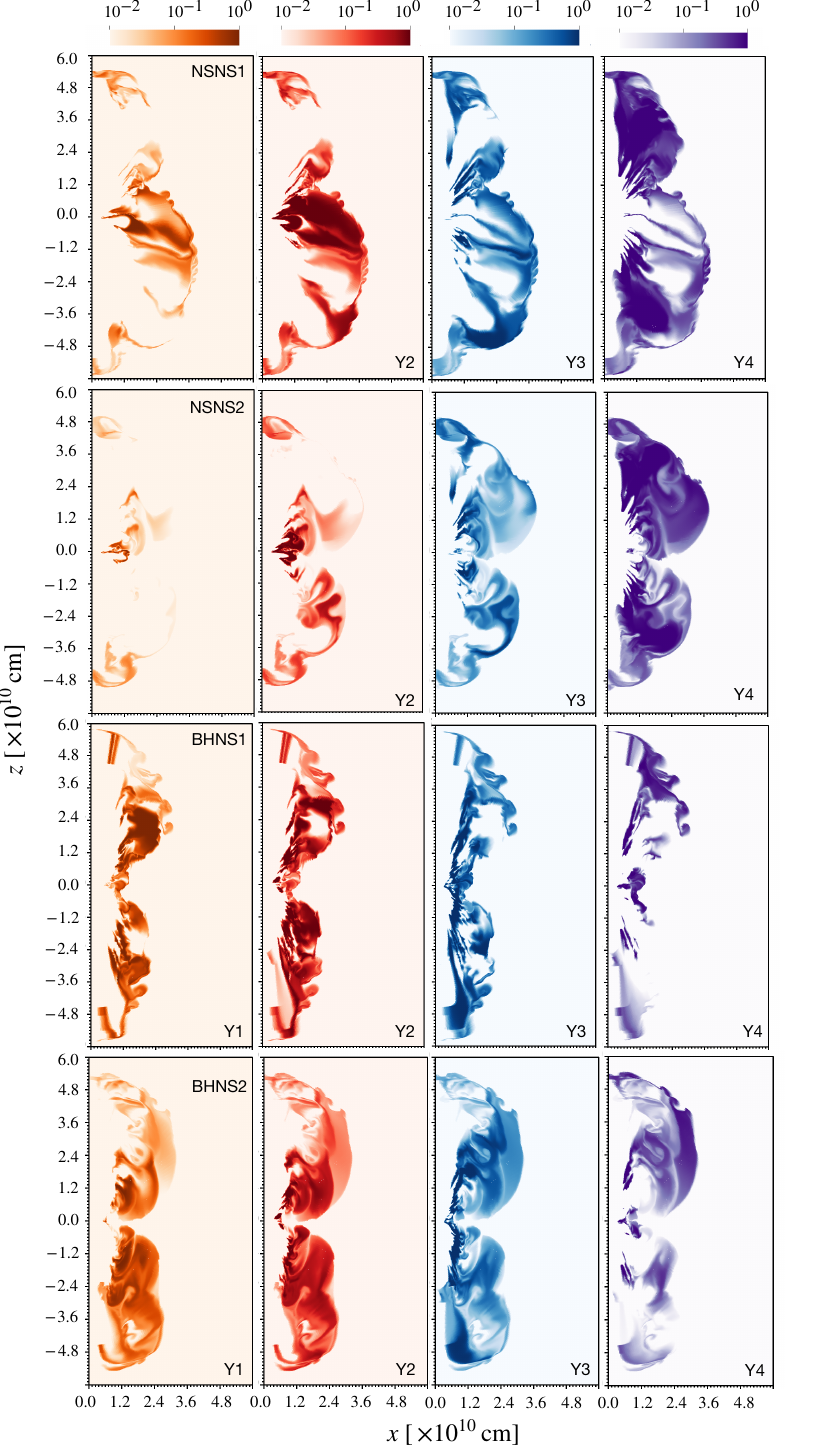}
    \caption{Components (Y1, Y2, Y3, Y4) of the disk wind tracked at $2~$s. These components are evolved as passive scalars, related with the electron fraction $Y_e$ as defined in eq.~\ref{eqn:compYe}.}
    \label{fig:distYe}
\end{figure*}

To follow a qualitative evolution of different components of the disk wind (Fig.~\ref{fig:histo_f}), we tracked the distribution of the ejected material using passive scalars\footnote{We track the chemical components using passive scalars $Yi$, which trace the geometrical evolution according to the advection of a mass fraction $Yi\,\rho$. \citet{PlewaMuller1999} demonstrated that the dynamics of the outflow is not altered adding of such passive scalars to the eqs.~\eqref{eqn:hydro_eqs} and \emph{Mezcal} code is able to follow the evolution of passive scalars. This technique is used to track the expansion of regions with different chemical compositions (see for example \citet{Suzuki2022}). However, it is important to note that no chemical evolution occurs.} associated with different ranges of electron fractions as 
\begin{equation}
    (Y1,Y2,Y3,Y4) = \begin{cases}
         (1,0,0,0) & {\rm if} \quad  0.1 <Y_e\leq 0.2,\\
         (0,1,0,0) & {\rm if} \quad 0.2 <Y_e\leq 0.3,\\
         (0,0,1,0) & {\rm if} \quad 0.3 <Y_e\leq 0.4,\\
         (0,0,0,1) & {\rm if} \quad 0.4 <Y_e\leq 0.5.
         \end{cases}
         \label{eqn:compYe}
\end{equation}
Every component $Yi$ represents the composition of the disk wind and evolves together with the fluid parcels. In Fig.~\ref{fig:distYe} at $t=2$~s, we show the tracked composition, where the most contrasting regions highlight the abundance of each component.

Rich lanthanide outflows can be categorized based on the electron fraction $Y_e \leq 0.25$ \citep[e.g.,][]{Metzger_2012,RosswogKorobkin2022}. In Fig.~\ref{fig:distYe}, components Y1 and Y2 represent qualitatively the segments characterized by a low electron fraction (the red kilonova component with high opacity), while components Y3 and Y4 represent segments with a higher electron fraction and low opacity (blue component).

In the case of NSNS1, this material accumulates closer to the equatorial region, whereas in NSNS2, it is confined to a smaller area. In contrast, the BHNS model exhibits a notably different distribution of high-opacity material, in an intermediate region between the poles and the equator. Both scenarios show that the presence or absence of DE strongly regulates the distribution of the chemical composition.

The initial distribution and mass of DE influence the final distribution of $Y_e$. This distribution is changed in our model only geometrically, and the chemical evolution is not included over time.


\subsection{Propagation of shock front}

\begin{figure}
    \centering
    \includegraphics[scale=0.35]{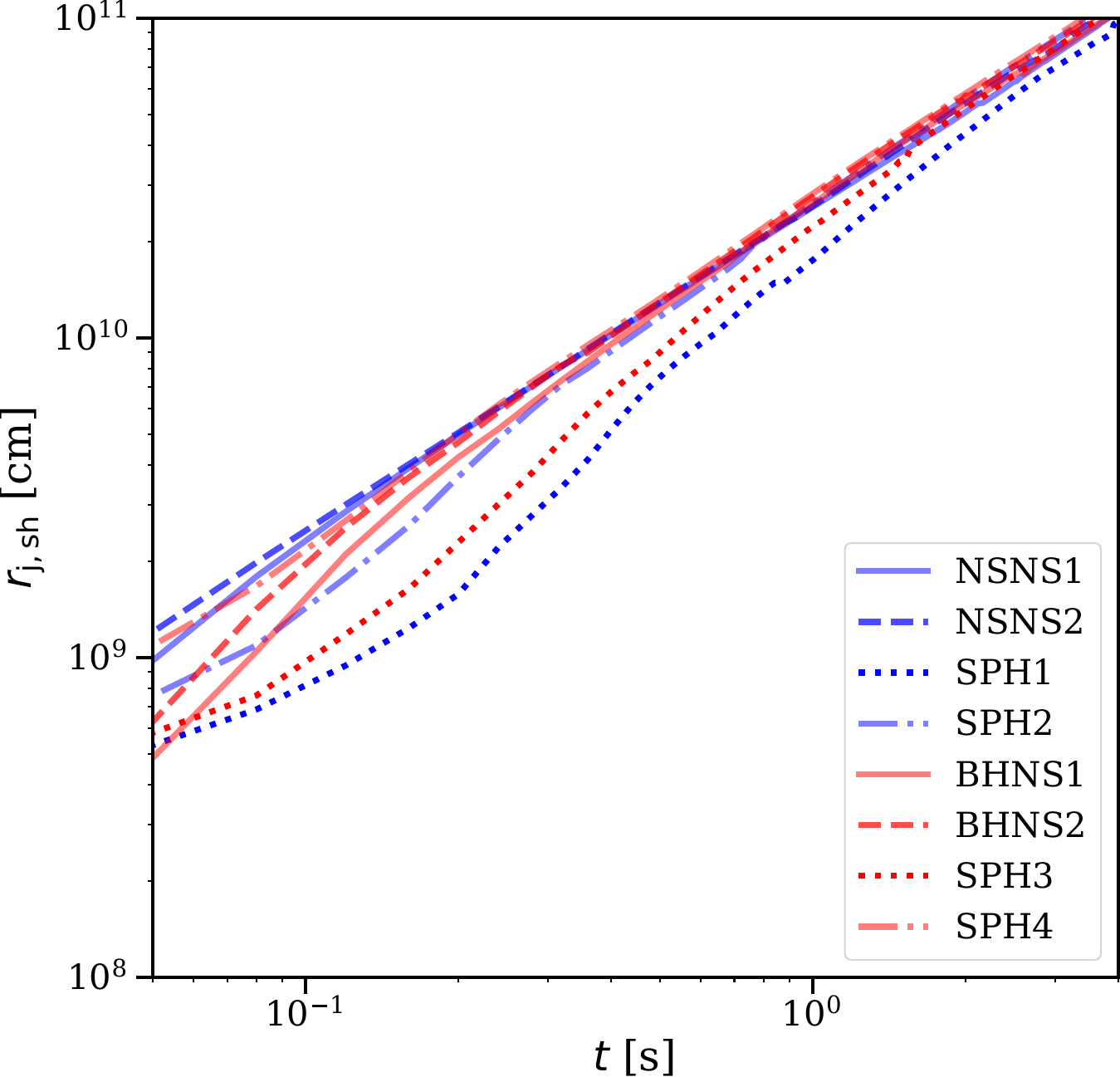}
    \caption{The evolution of the shock front position of the jet is presented on a logarithmic scale. The jet models associated with the NSNS merger scenario are denoted in blue, while red denotes the BHNS merger scenario.}
    \label{fig:shock_front}
\end{figure}

In Fig.~\ref{fig:shock_front} we show the propagation of the shock front $r_{\rm j,sh}$ for each jet listed in Table \ref{tab:list_models}. During the initial two seconds of evolution, collimated models can be distinguished by different slopes, since each undergoes different acceleration. However, at later times, the propagation slope becomes similar. In contrast, non-collimated jets, SPH1 and SPH3, which start with different values of $\Gamma_j$, exhibit a remarkable difference in their propagation evolution. This discrepancy is due to their jet heads not drilling efficiently into the atmosphere, resulting in a substantial accumulation of mass ahead of the jet.

\begin{figure*}
    \centering
    \includegraphics[scale=0.41]{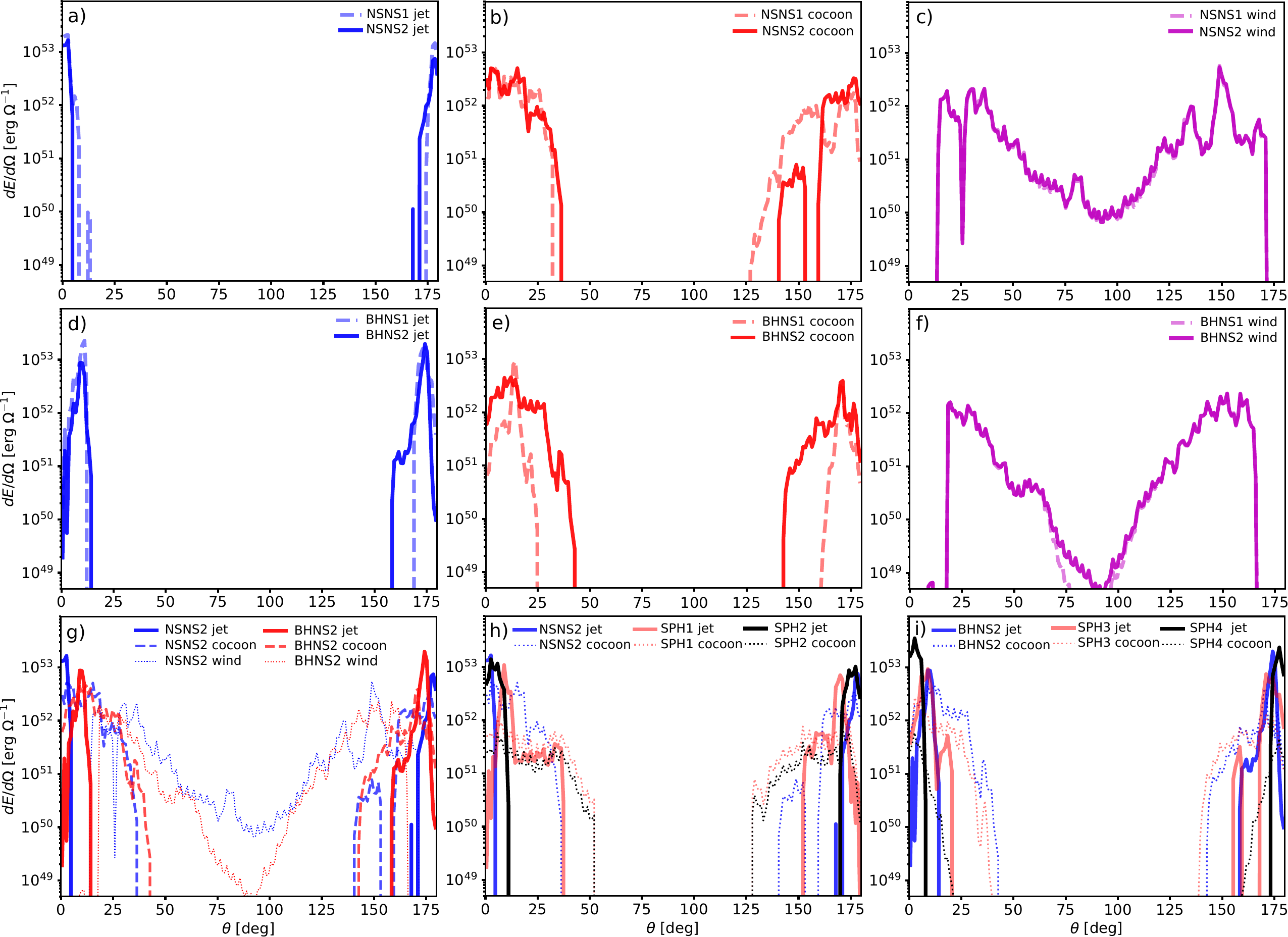}
    \caption{The energy distribution at $t=2~$s, estimated by eqs. \eqref{eqn:tot_en}-\eqref{eqn:tot_en_wind}. The energy within the jet is determined by the contribution from fluid exhibiting high Lorentz factors $\Gamma \geq \Gamma_j$. In contrast, the energy content of the cocoon derives from material moving within a range $\Gamma_j > \Gamma \geq 2$. Finally, the slower material contributes to the energy of the wind. In panels a) through c), we illustrate each of these energy contributions and a comparison between two distinct models for the NSNS scenario. A similar comparison is provided in panels d) through f) for the BHNS scenario. Panel g) shows a comparison between the NSNS2 and BHNS2 models. Panels h) and i) present the comparisons related to jet collimation and cocoon expansion.}
    \label{fig:ene_dist_sep}
\end{figure*}

Fig.~\ref{fig:ene_dist_sep} shows the energy structure of the jet, cocoon, and wind, determined by eqs.~\eqref{eqn:tot_en_jet}, \eqref{eqn:tot_en_cocoon}, and \eqref{eqn:tot_en_wind}, respectively. In panel a), we observe a similar structure for both jets, NSNS1 and NSNS2, with only a minor difference in collimation between them. A similar effect is seen in panel d) for models BHNS1 and BHNS2, except for the asymmetry at the south pole in model BHNS2, where the jet exhibits a notable extension of 10 degrees, larger compared to model BHNS1.

More significant differences and asymmetries can be observed in the energy distribution of the cocoon in panels b) and e). Panel b) shows a comparison between the energy distribution of the cocoon for models NSNS1 and NSNS2. The DE impacts the energy distribution in the south region of model NSNS2, but in general, the distribution is almost the same. A more notable difference is illustrated in panel e) for models BHNS1 and BHNS2, where the presence of DE affects the structure of the cocoon.

The energy distribution of the wind in panels c) and f) exhibits minimal discrepancies between the scenarios with and without DE. A distinctive contrast emerges when comparing the NSNS and BHNS scenarios, particularly in the equatorial region shown in panel g).

Panel g) provides a global illustration and shows all three energy components. Additionally, this panel offers a comparison between the two scenarios.

Panels h) and i) illustrate the effect of the disk wind on the evolution of the jet and cocoon. To explore this, we compare models NSNS2 and BHNS2 to jets injected into a spherical DE (SPH1-SPH4). In panel h), it is clear that the NSNS2 jet is collimated, and its cocoon exhibits less lateral spreading compared to the manually collimated jet in SPH2 and the non-collimated jet in SPH1. This suggests that the disk wind not only induces collimation but also restricts the lateral spreading of the cocoon.

On the other hand, in panel i), the BHNS2 jet shows similar lateral spreading to the non-collimated jet SPH3, while exceeding the lateral spreading observed in the collimated case SPH4. In this scenario, the disk wind exhibits a different sideways expansion, and the jet collimation is less efficient than the NSNS2 model.

\begin{figure*}
    \centering
    \includegraphics[scale=0.41]{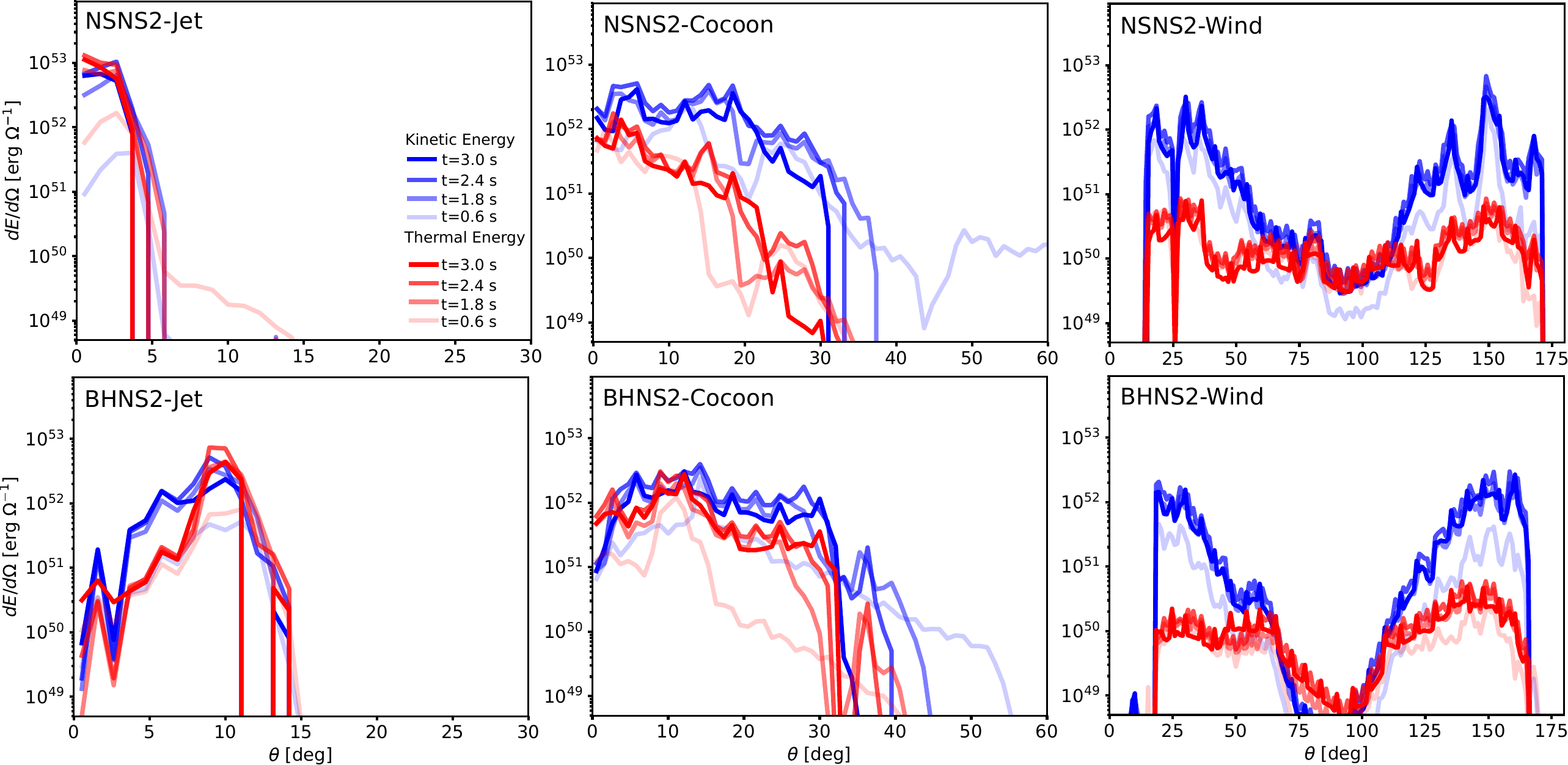}
     \caption{The energy evolution of each contribution for both the NSNS2 model (upper panel) and the BHNS2 model (lower panel). In this plot, thermal energy is represented by the colour red, while kinetic energy is depicted in blue, following the definitions as outlined in Eqs. \eqref{eqn:kin_en}, \eqref{eqn:th_en}, and \eqref{eqn:tot_en}-\eqref{eqn:tot_en_wind}.}
    \label{fig:ene_dist_sep2}
\end{figure*}

In Fig.~\ref{fig:ene_dist_sep2} we present the time evolution of the kinetic and thermal energy, separately for the jet, cocoon, and wind, in the models that contain DE. In the following paragraphs, we describe the evolution of each component.

The jet evolution proceeds as follows. At the beginning, the thermal energy dominates over the kinetic because the jet injection is pressure-dominated. At late times, kinetic and thermal energies have the same order of magnitude. The jet region does not present a complete conversion of the thermal to kinetic energy. 

The cocoon evolution is the following. The final shape (at late times) results from the continuous interaction with the jet, where the energy injection takes place close to the poles. In the case of the NSNS2 model, which presents a more collimated jet, the kinetic energy dominates over the thermal energy at late times, across the jet, because thermal energy is converted into kinetic energy. The model BHNS2 presents a similar distribution of thermal and kinetic energy, albeit there is a slower transfer of thermal energy to kinetic. 

In the polar region, the kinetic energy of the wind is more dominant than its thermal energy. This area contains the most external material within the system, interacting with a simple external medium. As a result, this material is freely expanding. On the other hand, at the equator, both kinetic and thermal components contribute equally, as materials from both poles collide, form shocks and provide subsequent heating. Here, the kinetic contribution is low due to the slower velocities of the wind in this region.

\section{Discussion}\label{sec:discussion}

%
The interaction between short GRB jets and their expanding dynamical ejecta significantly modifies the propagation and structure of these jets \citep[e.g.,][and references therein]{Salafia2023structuredjets}. Our results emphasize the critical need to distinguish between the components of the post-merger environments, namely the disk wind and DE (Fig.~\ref{fig:cartoon}). Each component exerts a unique influence on the propagation and structure of the jet. In this study, we focused on exploring the impact of the post-merger disk wind on the dynamics of short GRB jets at larger scales. The disk wind, as remapped in our simulations, results from the interplay between magnetized wind dynamics and neutrino cooling, presenting a more realistic stratification of the density and velocity distribution than usually considered \citep[e.g.,][]{Duffell2015,duffell18,Hamidani2020,Urrutia2021ShortGRBS,GottliebNakarExpandingMedia2021,Hamidani2023,Mpisketzis2024}. Moreover, our initial conditions consider pressure obtained from r-process nucleosynthesis. The nuclear pressure differs from that of the ideal gas and is contributed by the electron and positron degeneracy pressure. The Helmholtz free energy of particles is used to determine the pressure, specific entropy and internal energy forming a thermodynamically consistent equation of state \citep{TimmesArnett1999ApJS,Timmes_2000}. Under these conditions, the outflows are sites for nuclear reactions that are heating the material, facilitating nucleosynthesis within the medium. The wind pressure in our model is thus resulting from the nuclear reaction dynamics, shaping the resulting flow pattern.

In the post-merger scenario a broad range of temperatures is present \citep[e.g.,][]{Rosswog2023}. Unlike studies that are assuming the initial ideal cold wind, our approach is different because we are extracting a more self-consistent pressure from disk winds, leading to a more accurate representation of the system's thermodynamics.

\subsection{The influence of the disk-wind outflow in the jet propagation}

Our results highlight the primary elements influencing the jet propagation: 1) density in front of the jet head, 2) pressure exerted by the post-merger environment, and 3) the transition region between the jet and the environment.


The angular distribution of the disk wind plays an important role in the propagation of 
{both} the jet and cocoon. For example, \citet{Aloy2005, Murguia2017} show that toroidal winds expanding while the jet head interacts primarily with a low-density funnel. As a result, the cocoon expands rapidly sideways, unlike the typical spherical wind. In addition, the material accumulates in front of the jet and its density is regulated by the angular wind distribution.


The jet propagation is also sensitive to its initial opening angle $\theta_j$. The jet starts collimating if at the launching point the following criterion is satisfied:
\begin{equation}
\rho_j h_j \Gamma_j^2/\rho_{\rm EJ} < \theta_j^{-4/3},
\label{eqn:ref_crit}
\end{equation}
where $\rho_{\rm EJ}$ is the density of the ejecta. The jet remains collimated if the density keeps constant or is stratified \citep{Bromberg2011,Nagakura_2014}. Alternatively, the jet can be re-collimated if the environment is cold, uniform and homologously expanded \citep[e.g.,][]{Hamidani2020,GottliebNakarExpandingMedia2021}. The initial collimation of the jet determines if the bow shock will be slim or pronounced, allowing material to accumulate in front of the jet \citep[e.g.,][]{Aloy2005,Nagakura_2014,Hamidani2020,Nativi2020,Urrutia2021ShortGRBS,GottliebNakarExpandingMedia2021}. In our study, we can predict if the jet keeps collimated at distances greater than the launching point by verifying this criteria eq. \eqref{eqn:ref_crit}. However, it is possible only for jets propagating in the spherical homologous ejecta with density $\rho_{\rm EJ}=\rho_{\rm DE-2}$ (See Table \ref{tab:list_models}).  

\citet{Hamidani&IOKAexpanding, GottliebNakarExpandingMedia2021} modify the criterion eq. \eqref{eqn:ref_crit} by introducing a temporal dependence in $\rho_{\rm EJ}$ for homologously expanding ejecta. However, their main assumptions consider a cold gas spherically distributed. In our work, we cannot predict analytically the future collimation for our most complete models because they are composed of dynamical ejecta and disk wind that is not a cold gas. Furthermore, the mixture of both components does not undergo a homologous expansion.

The expansion of the cocoon is also modified by the interaction with the post-merger disk wind and the dynamical ejecta (Fig.~\ref{fig:ene_dist_sep}). The pressure and velocity distribution of the wind either expands or compresses the cocoon material, contributing as well to the final jet collimation.

Examining different initial jet opening angles in combination with post-merger environment scenarios reveals the influence of realistic disk wind ejecta on jet collimation. Specifically, this is observed in models transitioning from non-collimated to collimated states, such as NSNS1 and NSNS2. However, generalizing these findings based on assumptions about source properties may not hold universally, as is the case for BHNS1 and BHNS2, where the jet do not change significantly its initial opening angle  (See Table \ref{tab:final_collimation}).

\begin{table}
    \centering
    \begin{tabular}{c|c|c|c|}
    Model  & Initial &  &   \\
           & Collimation & $\theta_j$ & $\theta_{j,f}$ \\ \hline\hline
    NSNS1  & Not & $15^\circ$ &  $5.7^\circ$ \\
    NSNS2  & Not & $15^\circ$ &  $3.6^\circ$ \\
    SPH1   & Not & $15^\circ$ &  $16.3^\circ$\\
    SPH2   & Yes & $8^\circ$  &  $11.5^\circ$\\ \\
    BHNS1  & Not & $15^\circ$ &  $12^\circ$\\ 
    BHNS2  & Not & $15^\circ$ &  $13.2^\circ$\\
    SPH3   & Not & $15^\circ$ &  $13^\circ$\\
    SPH4   & Yes & $4.7^\circ$&  $7.8^\circ$\\ \hline
    \end{tabular}
    \caption{We present a list of our performed models along with the initial collimation, based on the criterion given in eq.~\eqref{eqn:ref_crit}. Additionally, we provide the specific values of the jet opening angle $\theta_j$ at the initial and $\theta_{j,f}$ final time of the jet’s evolution.}
    \label{tab:final_collimation}
\end{table}

\subsection{The effects of disk wind on the early jet structure}

The development of a structured jet is evident from the angular distribution observed in the Lorentz factor and luminosity during propagation at small scales \citep{Kathirgamaraju2019, James_2022, JaniukJames2022}. The jet structure can be partially preserved at intermediate scales if the interaction occurs in a low-density environment, such as $\dot{M}\sim 10^{-4}M_\odot\,$s$^{-1}$ \citep{Urrutia2021ShortGRBS} and lost in denser environments $\dot{M}\sim 10^{-2}M_\odot\,$s$^{-1}$ \citep{Nativi2022}. The post-merger environment potentially affects the jet structure due to the collimation criterion at the launching point \citep[e.g.,][]{urrutia22_3D} by modifying eq. \ref{eqn:ref_crit}), the mass-loss rate can be written as 
\begin{equation}
\dot{M}_w < {2 L_j v_w}/{r_j^2 \theta_j^{2/3}}\, .
\label{eqn:dot_structure}
\end{equation}
However, this estimation was made by assuming a spherical environment.

In this work, we did not consider the implementation of a structured jet at the initial conditions, since the average mass-loss rate estimated from GRMHD simulations from \citet{Nouri_2023} exceeds values $\dot{M}\gtrsim10^{-2}\,M_\odot$. Based on the results of \citet{Nagakura_2014,Nativi2022}, we suggest that any initial structure will not be preserved but will be modified drastically after interacting with the post-merger environment. 

Our work suggest that the time scales ($t\lesssim 1$~s) when the $r-$process is important, the post-merger winds are not in homologous expansion and in some cases influence the pressure balance between the jet and cocoon. However, the wind injected in our simulations is self-consistent with the $r-$process only for $t\lesssim 0.3$~s. For the remainder of the injection time we rely on other assumptions (see section \ref{sec:methods}). 

Further studies are needed to investigate the interaction between the jet and a postmerger environment with a self-consistent chemical evolution at least for times of the order of the mean lifetime of the central engine $t_j$, and to verify if the post-merger ejecta achieve homologous expansion before this time. It is essential to validate this within the known predictions on jet collimation and conservation of the initial jet structure.

\subsection{The evolution of energy distributions and implications to large scales}

Throughout our simulations, the dominance of thermal energy within the jet persists until late times, indicating that interactions with the cocoon are primarily driven by thermal energy transfer. The energy evolution of the cocoon is significantly influenced by the post-merger environment. For example, in model NSNS2 (Fig.~\ref{fig:ene_dist_sep2}), the interaction between the cocoon and disk wind is dominated by kinetic energy, suggesting that this region is shaped by the velocity distribution. On the other hand, model BHNS2 exhibits a thermalized cocoon at late times.

Homologous expansion, achieved only through complete thermal-to-kinetic energy conversion, occurs distinctly in the system. While the distribution of heavy wind material, particularly near the polar regions, undergoes dominant kinetic energy expansion, this expansion is non-uniform. The simple scenario assumes a homologous wind expansion, yet significant thermal energy retention occurs in the polar regions. Disruptions in the jet's kinetic energy conversion by the high-speed material in the equatorial regions may impact estimations of the jet emission, warranting careful modeling in the future kilonova studies \citep[e.g.,][]{Rosswog2017,Radice_2018,radice2018b,Cowan2021,Curtis2023,Arcones2023,Nouri_2023}.

In previous studies \citep[e.g.,][]{Aloy2005,Duffell2015,duffell18,Hamidani2020,Urrutia2021ShortGRBS,Murguia_Berthier_2021,Hamidani2023}, the long-term evolution of the cocoon results in its expansion, eventually engulfing the post-merger wind. This effect occurs because the velocity of the cocoon exceeds the uniform velocity of the homologous wind. However, in our scenarios, the initial wind material exhibits higher velocities near the jet region (Fig.~\ref{fig:histo_f}). Consequently, this faster wind material moves together with the cocoon, and the two elements become mixed (Fig.~\ref{fig:distYe}).

An important application of the jet and cocoon evolution at large scales involves extrapolating energy distributions to larger distances of $r\sim 10^{16}~$cm, in order to estimate the afterglow radiation \citep[e.g.,][]{duffell18,lazzati18,MooleyNature2018,Nathanail2020,Urrutia2021ShortGRBS,Dichiara_2021,Nativi2022,Pais_2023,Mpisketzis2024}. However, our results suggest the need to carefully consider the evolution of the energy distribution before such large distances, as the transformation from thermal to kinetic energy remains significant. Ideally, extrapolation should be done once all thermal energy has been converted to kinetic energy. This presents a potential application for modeling future observations of \emph{off-axis} GRBs and upcoming multi-wavelength surveys \citep[e.g.,][]{Salafiaetal2016,Ghirlanda2019,Gottlieb2019,Dichiara_2021}.

\section{Conclusions}\label{sec:conclusions}

We perform two-dimensional SRHD numerical simulations to study the large-scale interaction between short GRB jets with post-merger disk wind outflows. These winds were remapped from previous GRMHD simulations of post-merger black hole accretion disk evolution. In particular, we follow the evolution of jets and outflows arising from NSNS and BHNS configurations.

Our results reveal crucial differences compared to common models that use homologously expanding winds:

\begin{itemize}
    \item Since we considered the impact of the nuclear heating via the r-process on the initial wind pressure, we observed that it leads to significant alterations in both jet collimation and cocoon expansion rate.
    \item Density stratification and velocity distribution of the wind induces asymmetric jets. 
    \item The angular structure of the energy components (thermal and kinetic) for the jet, cocoon, and wind presents crucial differences compared to homologous models. This suggests that the GRB afterglow emission for \emph{off-axis} observers will be modified since the radiation estimation is strongly connected with the energy distribution.
    \item During the time evolution of the jet, thermal energy is not converted into kinetic energy. In the NSNS2 model, the cocoon exhibits thermal energy conversion at large angles, and this is preserved at the same order of magnitude as in the BHNS2 model. The kinetic energy dominates near the poles, while thermal energy dominates at the equator.
\end{itemize}

Notwithstanding, the energy structure and the expansion of the expelled material are consequences of the assumed configuration, namely either an NSNS or BHNS merger. The results are sensitive to the interaction between the jet, the disk wind, and the DE. This is demonstrated, for example, by tracking the evolution of the chemical composition of the disk-wind outflow. 

\begin{itemize}
    \item The DE induces a spreading of the disk wind. As a result, the interaction between the jet and the disk wind becomes more favorable.
    \item In the post-merger scenario of NSNS, the lanthanide-poor material (more transparent) tends to accumulate near the poles, while the more lanthanide-rich (opaque) material is found closer to the equator. However, in the case of BHNS, the distribution remains more mixed.
    \item The jet remains free of pollution for all models.
\end{itemize}

For this study, we did not consider the influence of magnetic fields on large-scale evolution. The presence of magnetic fields induces the collimation of the jet, making its propagation more stable, particularly when it is polluted by the progenitor environment \citep[e.g.,][]{Nathanail2020,gottlieb2022c,Pavan2023,Pais_2023,Pais_2024,Gottlieb2023-BHNS}. 

Our study emphasizes the critical influence of the post-merger disk wind on jet propagation. This element must be carefully managed since its main characteristics are sensitive to engine properties such as spin, magnetic field strength, and neutrino cooling. Since the disk wind's contribution can potentially affect the final jet structure, the properties of the central engine will be reflected in the estimation of electromagnetic counterparts.

Finally, although our disk wind is more realistic than homologous cases, the DE (expelled during the merger) used for the NSNS2 and BHNS2 models is a simplified version of a realistic case. The results would be improved if the realistic disk wind models were embedded within self-consistent post-merger environments. 

\appendix

\section{Resolution study}

The maximum resolution of our grid configuration is based on two criteria. The first criterion is finding a resolution similar to those reported in previous studies. This resolution is defined as the minimum resolved space at the smallest cell. For example, in spherical coordinates \citet{Pavan2023} report the maximum resolution at the smallest cells as $\Delta r \approx r_{\rm inj} \Delta \theta \approx 4.44 \times 10^{5}$~cm. However, this study is restricted to following the jet over relatively short distances with respect to the central engine, from $r_{\rm inj} \approx 3.8\times 10^7$~cm to $r\sim 6\times 10^8$~cm. \citet{Hamidani2020} have reported simulations following jets to large scales with two resolutions at the coarsest level $\Delta r_{\rm min} \sim 10^5$~cm and $\Delta r_{\rm min} \sim 10^6$~cm. In similar works, \citet{Murguia-Berthier2014} have reported a maximum resolution of $\Delta r_{\rm min}\approx 6.25\times 10^6$~cm. In summary, studies of jet interaction in the context of short GRBs, present a trust range of maximum resolution of $10^5-10^6$~cm. 

In our present study, we aim to examine the jet expansion at $r\sim 10^{11}$~cm from an injection radius of $r=10^8$~cm. This choice is based on the fact that it is the maximum distance at which our previous accretion simulations \citep{Nouri_2023} can report unbounded material. Then, our maximum resolution should be of the order of $\Delta r \approx r_{\rm inj} \Delta \theta \approx 10^{6}$cm.

\begin{figure}
    \centering
    \includegraphics[width=0.95\linewidth]{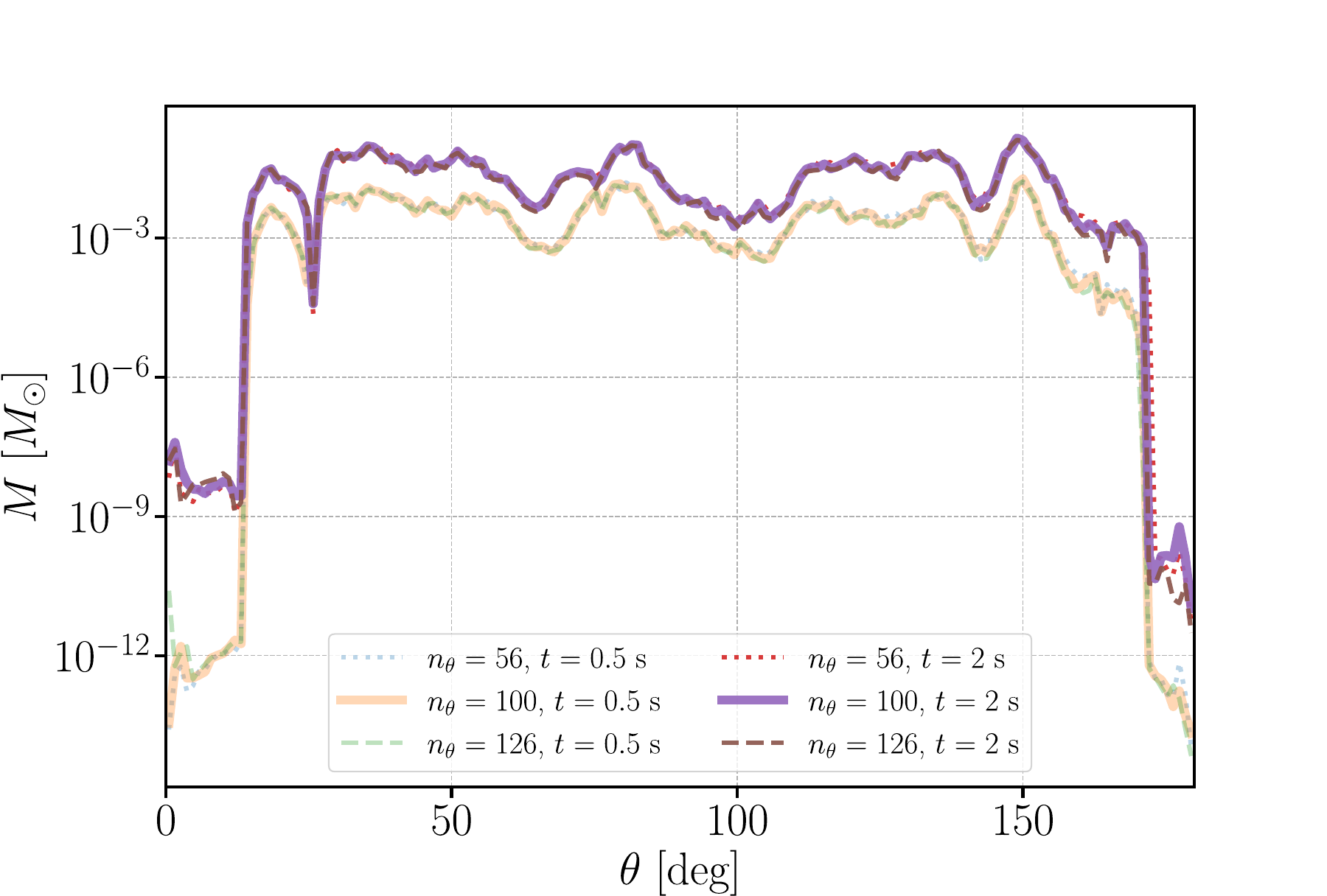}
    \includegraphics[width=0.95\linewidth]{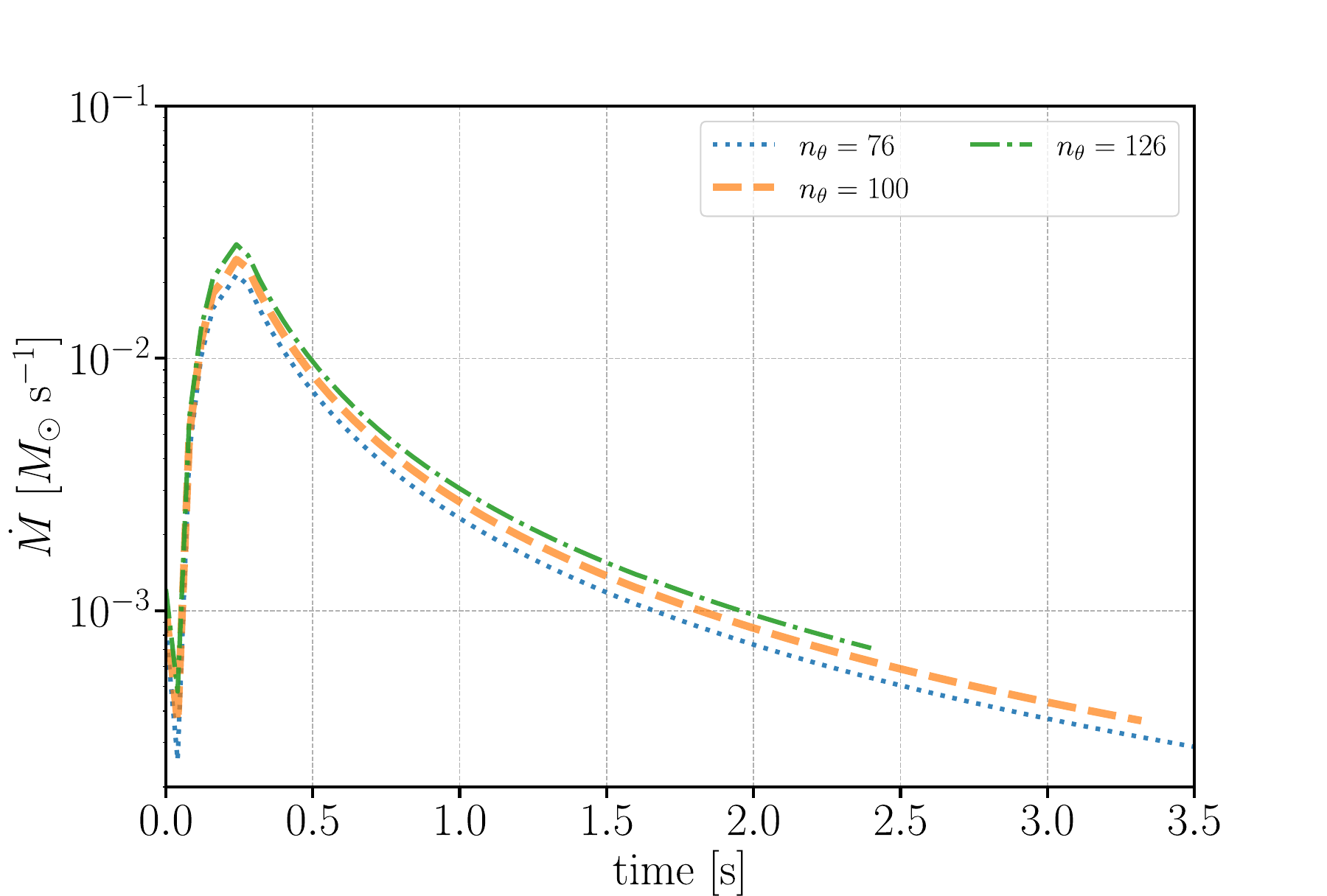}
    \includegraphics[width=0.95\linewidth]{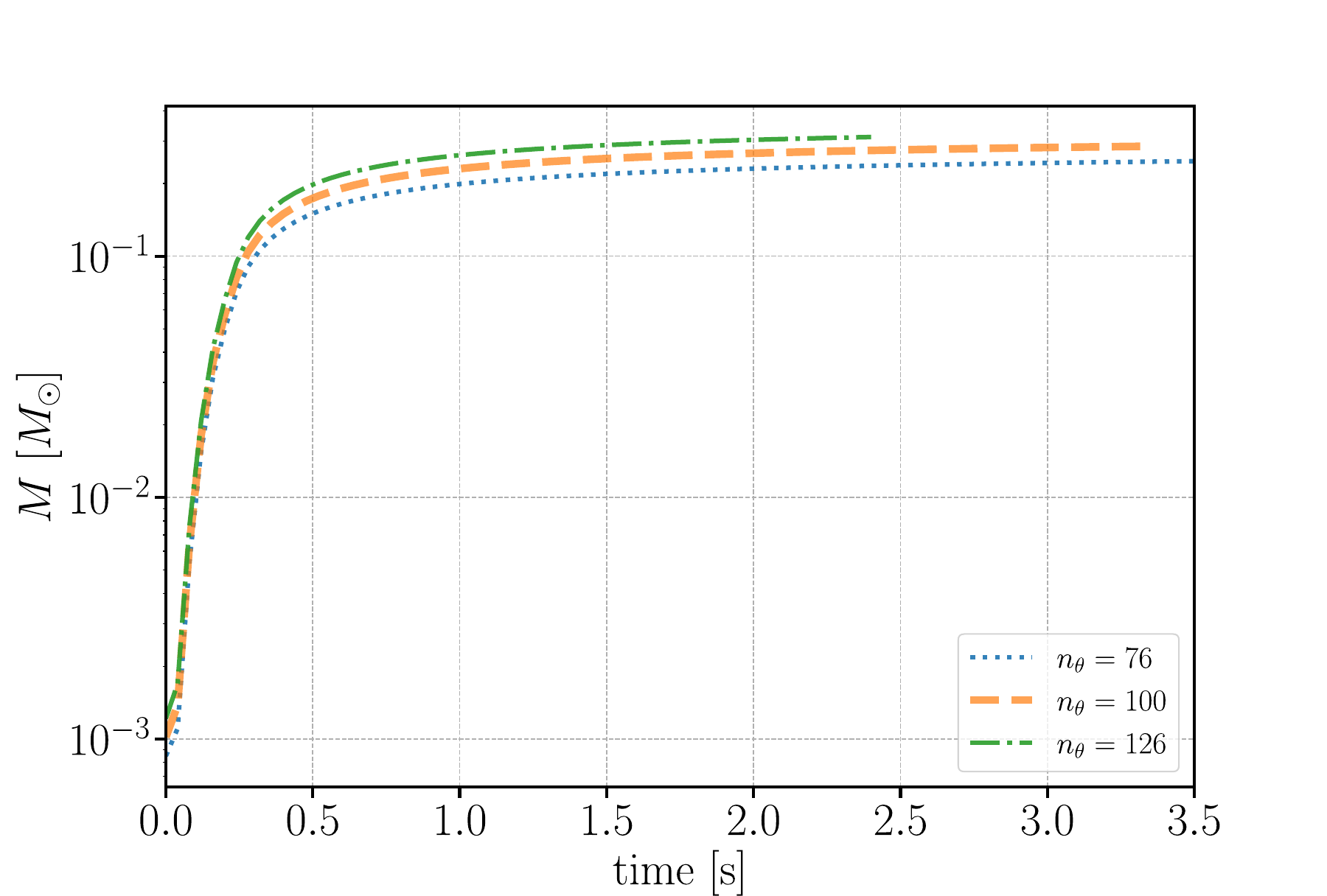}
    \includegraphics[width=0.95\linewidth]{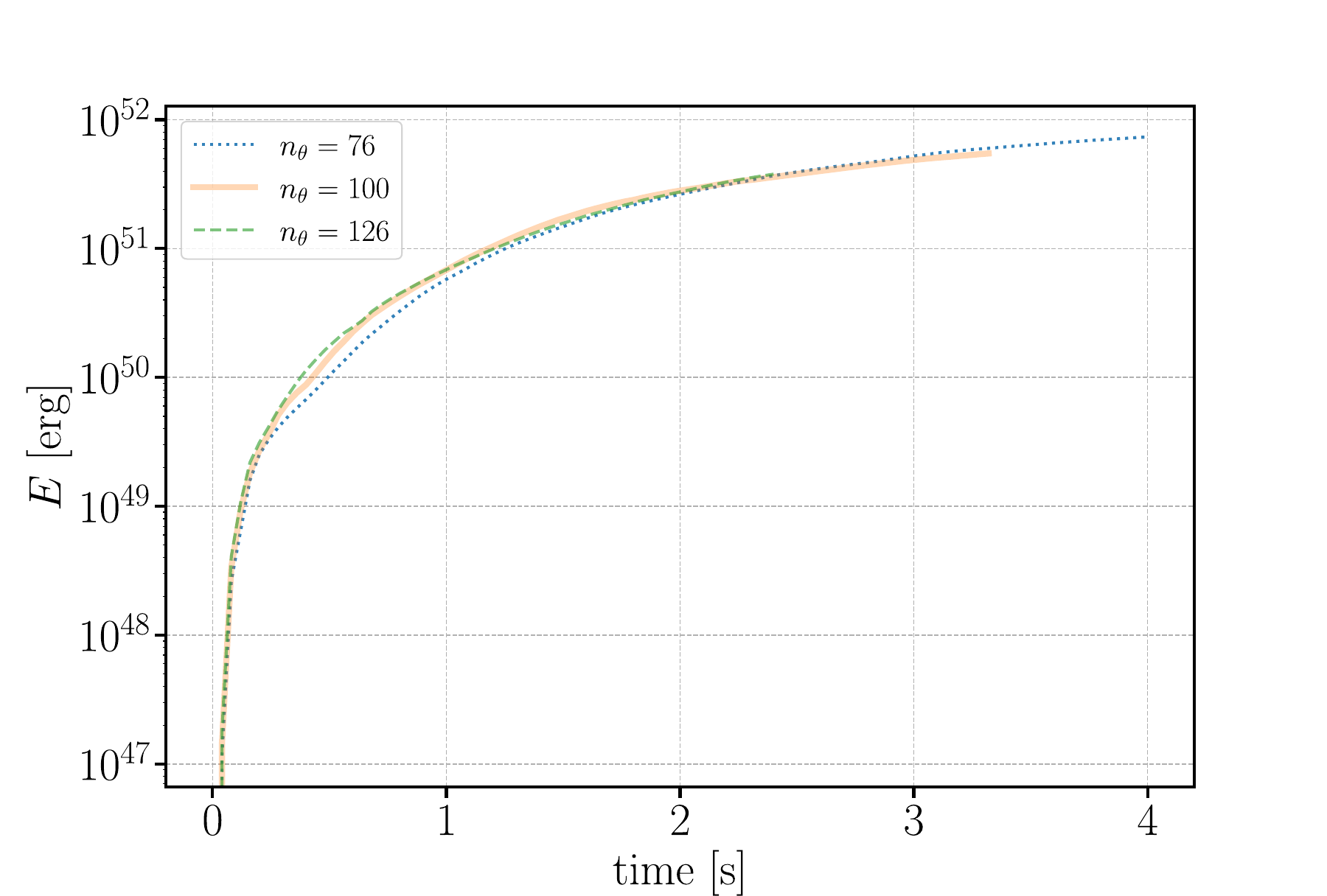}
    \caption{Estimations from model NSNS2 varying the resolution (or number $n_{\theta}$) in the $\theta$ direction. The top panel displays the angular distribution of the injected mass of the wind at two different times. The subsequent panels show the evolution of the mass loss rate $\dot{M}$ of the wind, the total injected mass $M$, and the total energy of the material whose Lorentz factor is $\Gamma \gtrsim 2$.}
    \label{fig:convergence_test}
\end{figure}

The second criterion is based on our resolution test. In our grid configuration, we are employing a number of levels of refinement of $n_l=4$, a number of cells $N_r=1\times 10^{4}$ along $r$ direction, and $N_\theta = 100$ in $\theta$ direction, we found a maximum resolution of $\Delta r \approx 1.49\times 10^6$~cm and $r_{\rm inj} \Delta \theta \approx 1.17 \times 10^{6}$~cm. It is the expected resolution of our adaptive mesh refinement and was based in $\Delta r= (r_{\rm max}-r_{\rm inj})/(N_r\cdot 2^{n_l -1})$ and $r_{\rm inj} \Delta \theta = r_{\rm inj}\theta_{\rm max}/(N_\theta \cdot 2^{n_l -1})$. Since, in this work, we are looking for angular distributions, its important to verify the resolution in $\theta$ direction, notwithstanding that in reality, our maximum number of cells is increasing according to $N_{\rm max}=N_\theta\cdot 2^{n_l -1}$. In Fig.~\ref{fig:convergence_test}, we present a convergence test, changing the initial configuration of the number of cells along the $\theta$ direction, as $N_\theta=76,\,100,\,126$. Based on such a figure, we opt to select the medium resolution of $N_\theta=100$, being a confidence value located in the middle of high and low resolution.

\section*{Acknowledgements}
We acknowledge the anonymous referee for a careful reading of the manuscript and for his/her constructive suggestions that improved it substantially. We thank Fabio De Colle, Diego L\'opez-C\'amara, H\'ector Olivares, Hiroki Nagakura and Leonardo Garc\'ia-Garc\'ia for useful discussions. This work was supported by the grant 2019/35/B/ST9/04000 from the Polish National Science Center. We gratefully acknowledge Polish high-performance computing infrastructure PLGrid (HPC Centers: ACK Cyfronet AGH) for providing computer facilities and support within computational grant no. PLG/2023/016178, PLG/2024/016972 and Warsaw Interdisciplinary Center for Mathematical Modeling. 

\section*{Data availability}

The data underlying this article will be shared on reasonable request to the corresponding author.

\bibliographystyle{mnras}
\bibliography{main} 

\bsp	
\label{lastpage}

\end{document}